\documentclass[12pt,letterpaper]{article}
\usepackage[utf8]{inputenc}

\usepackage{graphicx,array}
\usepackage{url}
\usepackage{color}
\usepackage{latexsym}
\usepackage{amsthm}
\usepackage{amsmath}
\usepackage{amssymb}
\usepackage{amsfonts}
\usepackage[numbers,sort&compress]{natbib}
\usepackage{bm}
\usepackage{bbm}
\usepackage{slashed}
\usepackage{mathrsfs}
\usepackage{enumerate}
\usepackage{tikz}
\usepackage{siunitx}
\usepackage{mdframed}
\usepackage{setspace}  
\usepackage{esvect}
\usepackage{physics}

\usepackage{tcolorbox}%

\usepackage{hyperref} 
\hypersetup{
    colorlinks=true,       
    linkcolor=red,          
    citecolor=blue,        
    filecolor=magenta,      
    urlcolor=blue           
}
\usepackage[all]{hypcap} 
\usepackage{multirow}
\usepackage{multicol}

\usepackage{natbib}
\usepackage[normalem]{ulem} 

\newcommand{\nc}{\newcommand}

\nc{\beq}{\begin{equation}}
\nc{\eeq}{\end{equation}}
\nc{\beqa}{\begin{eqnarray}}  
\nc{\eeqa}{\end{eqnarray}}  
\nc{\bit}{\begin{itemize}}  
\nc{\eit}{\end{itemize}}  

\newcommand{\ie}{{\it i.e.}}
\newcommand{\Mpl}{M_{\rm pl}}

\nc{\Trhi}{T_{\rm rh}^0}
\nc{\Tform}{T_{\rm form}}
\nc{\Tscaling}{T_{\rm scal}}
\nc{\Tdomin}{T_{\rm dom}}
\nc{\Tcaus}{T_{\rm caus}}
\nc{\TQCDj}{T_{\rm QCD}^j} 
\nc{\Tann}{T_{\rm ann}}
\nc{\Trhw}{T_{\rm rh}^{\rm w}}
\nc{\mpl}{M_{\rm pl}}

\usepackage{floatrow}
\newfloatcommand{capbtabbox}{table}[][\FBwidth]

\usepackage{blindtext}

\title{ 
 {\bf QCD-Collapsed Domain Walls:
 }\\
 {\bf \large QCD Phase Transition and Gravitational Wave Spectroscopy}
\author{\large Yang Bai, Ting-Kuo Chen, and Mrunal Korwar}
\date{\small \it 
Department of Physics, University of Wisconsin-Madison, Madison, WI 53706, USA
}
}

\begin{document}

\maketitle

\setlength{\parskip}{0.2ex}

\begin{abstract}
For a discrete symmetry that is anomalous under QCD, the domain walls produced in the early universe from its spontaneous breaking can naturally annihilate due to QCD instanton effects. The gravitational waves generated from wall annihilation have their amplitude and frequency determined by both the discrete symmetry breaking scale and the QCD scale. The evidence of stochastic gravitational waves at nanohertz observed by pulsar timing array experiments suggests that the discrete-symmetry-breaking scale is around 100 TeV, assuming the domain-wall explanation. The annihilation temperature is about 100 MeV, which could naturally be below the QCD phase transition temperature. We point out that the QCD phase transition within some domains with an effective large QCD $\theta$ angle could be a first-order one. To derive the phase diagram in $\theta$ and temperature, we adopt a phenomenological linear sigma model with three quark flavors. The domain-wall explanation for the NANOGrav, EPTA, PPTA and CPTA results hints at a first-order QCD phase transition, which predicts additional gravitational waves at higher frequencies. If the initial formation of domain walls is also a first-order process, this class of domain-wall models predicts an interesting gravitational wave spectroscopy with frequencies spanning more than ten orders of magnitude, from nanohertz to 100 Hz.  
\end{abstract}

\thispagestyle{empty}  
\newpage    
\setcounter{page}{1}  

\begingroup
\hypersetup{linkcolor=black,linktocpage}
\tableofcontents
\endgroup

\newpage

\section{Introduction}
It has been pointed out a half century ago that spontaneous breaking of discrete symmetries can lead to productions of domain walls in the early universe and overclose the universe for a high symmetry-breaking scale~\cite{Zeldovich:1974uw}. Additional ad-hoc explicitly symmetry breaking operators are usually introduced to bias the potential energy in different domains and collapse walls. One more natural way to collapse the domain walls is to use the known QCD instanton effects in the Standard Model (SM), which was pointed out in Ref.~\cite{Preskill:1991kd}. The minimal assumption is that the discrete symmetry is anomalous under the QCD interactions such that the QCD instanton effects generate an effective potential for the discrete-symmetry order parameter, explicitly break the symmetry and collapse the walls at around the temperature of QCD phase transition.  

Discrete symmetries are ubiquitous among particle physics models. For instance, the spontaneously breaking of time-reversal was studied by T. D. Lee a long time ago~\cite{Lee:1973iz}. More recently, discrete matter symmetries are introduced in the Nelson-Barr mechanism~\cite{Nelson:1983zb,Nelson:1984hg,Barr:1984qx} to solve the strong CP problem~\cite{tHooft:1976rip,Baker:2006ts}. To explain quark and lepton masses and mixings, many discrete Abelian and non-Abelian flavor symmetries have also been introduced to achieve certain matrix structures (see \cite{Everett:2008et} for instance). Some of those discrete symmetries could be anomalous under the QCD interactions. Early literature about general discrete symmetry anomalies can be found in Refs.~\cite{Ibanez:1991hv,Araki:2008ek}, while more specific studies related to flavor symmetries can be found in Refs.~\cite{Chigusa:2018hhl,Gelmini:2020bqg}.

In this paper, we want to point out an interesting feature about the QCD-anomalous discrete symmetries. The QCD dynamics inside different domains can be thought of as different QCD's with different $\theta$ angles (the strong CP angles). The domains with $\theta = 0$ are energetically preferred if a Nelson-Barr-like mechanism exists to enforce zero $\theta$ angle before the discrete symmetry breaking (the QCD-anomalous discrete symmetry could be embedded in the Nelson-Barr mechanism).  The finite-temperature QCD phase transition for $\theta=0$ is a crossover one~\cite{Aoki:2006we,Bhattacharya:2014ara}. However, the QCD phase transition with a $\theta$ angle close to $\pi$ could be a first-order one. One hint about this possibility is the first-order phase transition along the $\theta$ direction at $\theta = \pi$~\cite{Dashen:1970et,Witten:1980sp,Gaiotto:2017tne,Gaiotto:2017yup}. Another hint is based on the phenomenological linear sigma model coupled to quarks (LSM$q$)~\cite{Pisarski:1996ne}, in which the finite-temperature QCD phase transition is shown to be a first-order one at $\theta=\pi$~\cite{Mizher:2008hf,Boomsma:2009eh}. Although a robust answer for the phase diagram in $\theta-T$ requires some non-perturbative tool, we will extend the two-quark-flavor study in Ref.~\cite{Mizher:2008hf} to the more realistic three-quark-flavor case and demonstrate a region of $\theta$ in $[\theta_c, 2\pi - \theta_c]$ centered at $\pi$ to have a first-order QCD phase transition. The existence of first-order QCD phase transition could have many phenomenological consequences including the formation of quark nuggets~\cite{Witten:1984rs, Oaknin:2003uv, Liang:2016tqc,Bai:2018vik}.

On the gravitational wave (GW) side, the annihilation of domain walls can generate stochastic gravitational wave background (SGWB) with its amplitude and frequency determined by both the discrete-symmetry-breaking scale $f$ or the wall tension $\sigma \sim f^3$ as well as the so-called potential bias parameter that measures the potential energy differences between different domains~\cite{Saikawa:2017hiv}. The models with QCD-anomalous discrete symmetries are more economical or predictive because the QCD instanton effects are the source of the potential bias parameter and have the known contributions at the QCD scale $\mathcal{O}(100\,\mbox{MeV})$. In our study here, we will use $\mathbb{Z}_N$ as an example to study the domain-wall evolution as well as the annihilation-generated GW. Other than studying the scaling case for domain wall evolution in the radiation-dominated universe~\cite{Hiramatsu:2010yn,Hiramatsu:2013qaa}, we also consider the case with a domain-wall-dominated universe and the corresponding GW. In our study, we pay some special attention to the cluster of potential bias values when $N > 2$ such that different domain walls annihilate at different temperatures and could generate a GW spectrum different from the minimal one with $N=2$ and one single annihilation temperature. Some of the GW spectrum features can be used to identify the group theory properties of the discrete symmetry and the detailed effective potential in terms of the domain-wall order parameter. 
 
On the GW experimental side and in the last few years, three of the current pulsar timing array (PTA) experiments NANOGrav~\cite{NANOGrav:2020bcs}, EPTA~\cite{Chen:2021rqp}, and PPTA~\cite{Goncharov:2021oub} had reported strong evidence for a common-spectrum red process across pulsars in their data. Their results were also confirmed by the IPTA collaboration~\cite{Perera:2019sca}, combining the data sets of the three collaborations. More recently, an evidence of the key Hellings-Downs correlation~\cite{Hellings:1983fr} to confirm the gravitational wave origin of the red spectrum has been shown at around $3\sigma$ confidence level by the NANOGrav collaboration~\cite{NANOGrav:2023gor}, which is also supported by EPTA~\cite{Antoniadis:2023ott}, PPTA~\cite{Reardon:2023gzh} and CPTA~\cite{Xu:2023wog}. Other than the explanation of astrophysical super massive black-hole binaries, many new physics model explanations have also been proposed including domain-wall collapse~\cite{Bian:2020urb, Chiang:2020aui, Ferreira:2022zzo, Wu:2022tpe, Bian:2022qbh, An:2023idh, King:2023cgv, Madge:2023cak}. 

The NANOGrav collaboration has also presented their results for the domain-wall explanation of GW sources~\cite{NANOGrav:2023hvm}. For the domain walls with the discrete symmetry anomalous under QCD, the preferred symmetry breaking scale $f$ is  around $100$~TeV with the annihilation temperature around 100 MeV, which is close to but below the QCD phase transition temperature. According to the phase diagram calculated in this paper based on the LSM$q$ model, some order-one fraction of domains have a first-order QCD finite-temperature phase transition. This means that the PTA data implies an important consequence of the strong dynamics during early-universe evolution, which could also have many other phenomenological consequences. 

Our paper is organized as follows. In Section~\ref{sec:model}, we introduce a discrete $\mathbb{Z}_N$ symmetry that is anomalous under QCD. Section~\ref{sec:evolution} contains the domain wall evolution of both the scaling behavior and domain-wall-dominated cases. The QCD phase diagram in $\theta - T$ is presented in Section~\ref{sec:phase-transition} with the detailed calculation in Appendix~\ref{sec:appendix}. The gravitational wave signatures from domain-wall annihilation, QCD and discrete-symmetry phase transitions are calculated in Section~\ref{sec:GW}. Some formulas for GW spectra from phase transition are kept in Appendix~\ref{sec:PTfomula}. 

\section{QCD-anomalous discrete symmetry}
\label{sec:model}
For simplicity, we consider the domain walls related to the spontaneous breaking of a $\mathbb{Z}_N$ symmetry. Introducing a complex scalar field $S$ transforming as $S \rightarrow e^{i\,2\pi/N} S$ under  $\mathbb{Z}_N$, a $\mathbb{Z}_N$-invariant (non-)renormalizable potential, $V(S)$, exists to determine the $N$-fold vacua after the symmetry breaking
\beqa
\langle S \rangle_j = f\, e^{i\,2\pi\,j/N} \,, \quad \mbox{with}\, \quad j = 0, 1, \cdots, N-1 ~. 
\eeqa
The discrete symmetry breaking scale $f$ will be assumed to be much higher than the QCD or the electroweak scale in this study.  For the simplest $\mathbb{Z}_2$ case, the order-parameter field could be a real scalar field with a simple renormalizable potential $V(S)= \frac{\lambda}{4} (S^2 - f^2)^2$. For $N > 2$, the single-field potential could be $V(S) = - m^2 SS^\dagger + \lambda (SS^\dagger)^2 - \mu (S^N + S^{\dagger N})$ with the mass dimension of $\mu$ as $4-N$. For $N > 4$, this potential contains non-renormalizable terms and could be replaced by a renormalizable model with additional fields charged under $\mathbb{Z}_N$. We note that while there exist a simple analytic relation between $f$ and the potential parameters $m^2,\lambda,\mu$ for $N\leq 4$, this is in general not the case for $N>4$. Therefore, we choose $f$ as an independent scale parameter for the rest of the study assuming that it can be determined from minimizing $V(S)$, since $f$ is more directly relevant to phase transition and GW physics. Furthermore, for odd $N$'s greater than 4, $V(S)$ is apparently unbounded from below, while for even $N$'s this can also be the case within specific parameter space, which implies the need for a UV-completion of the model. For the purpose of this study, we assume that the UV-completed model exists and does not affect the phenomenological study here.

Parametrizing $S=\vert S\vert e^{i\theta}$, the angular field has the following effective Lagrangian
\begin{equation}\label{eq:Lagrangian:angular}
    \mathcal{L}_\theta = f^2\,\partial_\mu\theta\,\partial^\mu\theta + 2\,\mu\,f^N\cos(N\theta) ~,
\end{equation}
which explicitly manifests the $\mathbb{Z}_N$ symmetry in the cosine potential with minima at $\theta=2\pi j/N$, with $j = 0, 1, \cdots, N-1$. For the class of models with the $\mathbb{Z}_N$ symmetry anomalous under the QCD interaction, the effective interaction below the scale $f$ but above the QCD scale is
\beqa
\mathcal{L} \supset -\frac{1}{32\pi^2}\,G^{\mu\nu}\widetilde{G}_{\mu\nu}\, \left[\theta_0 + \sum_\psi  2\,q_{\psi}\,C(\bm{r}_{\psi})\,\theta \right] ~. \label{eq:Leff}
\eeqa
Here, $\widetilde{G}_{\mu\nu}=\frac{1}{2}\epsilon_{\mu\nu\alpha\beta}G^{\alpha\beta}$ with $G^{\alpha\beta}$ as the gluon field tensor; $\theta_0$ is the UV QCD $\theta$ angle; $C(\bm{r}_{\psi})$ is the Dynkin index for the representation of a chiral fermion $\psi$ under $SU(3)_c$ with $C(\bm{3}) = 1/2$;  $q_{\psi}$ is a (mod $N$) integer and is the $\mathbb{Z}_N$ charge of the heavy chiral fermion $\psi$ that obtains a mass after $\mathbb{Z}_N$ breaking. Note that, in different domains with different $\theta=\langle S \rangle_j$, the effective QCD $\theta$ angle is $\theta_j = \theta_0 + \sum_\psi 2\,q_{\psi}\,C(\bm{r}_{\psi}) 2\pi j/ N$. For $\theta_0 = 0$ and $n_f$ number of $\psi$ fermions with $q_\psi = 1$ and in $\bm{3}$ of $SU(3)_c$, one has $\theta_j =  2\pi\,j\,n_f/N$.  To have all domain walls collapsed by the QCD instanton effects or distinct $\theta$ angles for different domain numbers $j$'s, a necessary condition is $\gcd{(n_f, N)} = 1$, where $\gcd$ stands for ``greatest common divisor''. Otherwise, the discrete symmetry is only broken to $\mathbb{Z}_{\gcd{(n_f, N)}}$ by QCD with remaining uncollapsed domain walls affecting Big Bang nucleosynthesis (BBN) observables. In the following, we will simply take $\gcd{(n_f, N)} = 1$ with QCD breaking all $\mathbb{Z}_{N}$ symmetry such that different domains have $\theta$ angles as $\theta_j = 2\pi\,j/N$ with $j = 0, 1, \cdots, N-1$.  We assume $\theta_0=0$ for the remainder of the study~\footnote{We anticipate some Nelson-Barr like models to solve the strong CP problem, as the Peccei-Quinn model~\cite{Peccei:1977hh,Peccei:1977ur} together with the discrete symmetry still has the domain-wall problem~\cite{Preskill:1991kd}.} and will discuss the related strong CP problem in Section~\ref{sec:discussion}.

After QCD phase transition, the QCD instanton effects generate an effective potential for $\theta$, biasing different domains. At $T=0$ and the leading order in chiral expansion, the two-flavor potential is~\cite{DiVecchia:1980yfw}
\beqa\label{eq:thetapotential}
V(\theta) &=& - m_\pi^2 f_\pi^2 \sqrt{1 - \frac{4\,m_u\,m_d}{(m_u + m_d)^2}\,\sin^2\left(\frac{\theta}{2}\right)}~.
\eeqa
Note that the above formula is valid in the small effective $\theta$ angle limit. For a large $\theta$ angle, the QCD vacuum deviates dramatically from the one used by the chiral Lagrangian (for instance, the operator $G_{\mu\nu}\widetilde{G}^{\mu\nu}$ may develop a large vacuum expectation value, which is absent in the ordinary chiral Lagrangian vacuum). Although a reliable effective potential in $\theta$ requires a non-perturbative derivation and is absent at the current moment, we will use the above potential to guide us through the qualitative evolution of domain walls. As a comparison, we also present the effective potential in the LSM$q$ model in Appendix~\ref{sec:appendix}. 

The total potential for $\theta$, combining Eqs.~\eqref{eq:Lagrangian:angular} and \eqref{eq:thetapotential} is 
\beqa\label{eq:totalthetapotential}
V_{\rm tot}(\theta) &=& -2\,\mu\, f^{N} \cos(N\,\theta)- m_\pi^2 f_\pi^2 \sqrt{1 - \frac{4\,m_u\,m_d}{(m_u + m_d)^2}\,\sin^2\left(\frac{\theta}{2}\right)}~,
\eeqa
where the first term respects the discrete $\mathbb{Z}_N$ symmetry, while the second term breaks the symmetry and acts as a bias term among different domains.

Taking the quark mass ratio $z \equiv m_u/m_d = 0.49$~\cite{Fodor:2016bgu} in the $\overline{\mbox{MS}}$ scheme at 2 GeV, $m_\pi = 135$~MeV and $f_\pi=92$~MeV, the maximum potential difference among different domains is 
\beqa
V_{\rm bias}^{\rm max} = 0.66\,m_\pi^2\,f_\pi^2 \approx (100.4\,\mbox{MeV})^4 ~\qquad \mbox{for}\,\quad N=\mbox{even} ~,\label{eq:potential}
\eeqa
and a smaller $N$-dependent $V_{\rm bias}^{\rm max}$ when $N = \mbox{odd}$. Note that the above effective potential provides the leading-order value of the topological susceptibility $\chi_{\rm {top}}^{1/4} = 76.4$~MeV and is close to the NLO~\cite{GrillidiCortona:2015jxo} and Lattice QCD results~\cite{Borsanyi:2016ksw}. 

Because the $j=0$ or $\theta_j = 0$ domain has the lowest effective potential, other domains with a nonzero $j$ will eventually disappear with the corresponding walls collapsing at a temperature (potentially) below the QCD phase transition temperature. Different walls could have different biased potential on their two sides and therefore could collapse at slightly different temperatures, which depend on $N$ and the detailed effective potential. 

In light of the similarity between the $\mathbb{Z}_N$-symmetric model in this study and the KSVZ axion model~\cite{Kim:1979if,Shifman:1979if}, we briefly compare the two models before concluding this section. Both models introduce a pair of heavy vector-like quarks that couple to the angular modes ($\theta$ and axion, respectively). For the $\mathbb{Z}_N$-symmetric model, the ``shift symmetry" of the angular mode is explicitly broken at a high scale $\sim f$, much above the QCD scale [see Eq.~\eqref{eq:Lagrangian:angular}], while for the QCD axion model the axion particle has its mass come from the QCD instanton effects and related to the QCD scale [similar to Eq.~\eqref{eq:thetapotential} with $\theta$ replaced by $\theta$ over the number of fermions]. For the $\mathbb{Z}_N$-symmetric model, the combination of the high-scale potential term and the QCD-instanton-generated potential term explicitly break the $\mathbb{Z}_N$ and hence collapse the domain walls. On the other hand, for the QCD axion model, a discrete subgroup of the $U(1)_{\rm PQ}$ symmetry is respected by the QCD instanton effects, which leads to the axion domain wall problem, unless one adds additional bias terms by hand~\cite{Sikivie:1982qv,Chang:1998tb,Sikivie:2006ni}.


\section{Domain wall evolution}
\label{sec:evolution}
The time-independent domain wall solutions are related to the topological structure of the symmetry breaking. For the simplest $\mathbb{Z}_2$ case, an analytic solution exists for the domain wall profile $S(z) = f \,\tanh[\sqrt{\frac{\lambda}{2}}\,f\,z]$ with the wall thickness as $\sim (\sqrt{\lambda} f)^{-1}$. The surface energy density or the wall tension is $\sigma \equiv \int^\infty_{-\infty} T_{00} dz = \frac{2\sqrt{2}}{3}\sqrt{\lambda}f^3$, which we will treat as a model parameter. For general $\mathbb{Z}_N$ case with $N>2$, the domain wall solution interpolating between the $\theta= 2\pi j/ N$ minimum at $z=-\infty$ and the $\theta= 2\pi (j+1)/ N$ minimum at $z=+ \infty$ is  
\begin{equation}\label{eq:dwprofile}
    \theta(z) = \frac{2\pi j}{N} + \frac{4}{N}\arctan{\left[\exp((N \mu f^{N-2})^{1/2}z)\right]}~\, ,
\end{equation}
and the domain wall tension is given by $\sigma = 16 \mu^{1/2}\,f^{N/2 +1}\,N^{-3/2}$ ~\cite{Saikawa:2017hiv}. Note that for $\mu \sim f^{4-N}$, we have $\sigma \sim f^3$. On the other hand, for $\mu \ll f^{4-N}$ one has $\sigma \ll f^3$ and a much lighter angular mode than the continuous symmetry breaking scale $f$ [see Eq.~\eqref{eq:Lagrangian:angular}], which is similar to the QCD axion case.
In the following study, we will consider a high inflation scale with the reheating temperature $\Trhi$ higher than the symmetry breaking scale or $\Trhi  \gg f$. A thermal phase transition for the discrete symmetry happens at a temperature $\Tform \sim f$ to form the domain walls via the Kibble-Zurek mechanism~\cite{Kibble:1976sj,Zurek:1985qw}. Therefore, we take $\sigma$ and $\Tform$ as two model parameters for both the $\mathbb{Z}_2$ and the general $\mathbb{Z}_N$ cases.  

\subsection{Overview of domain wall evolution}
\begin{figure}[th!]
\centering
    \includegraphics[width=0.48\textwidth]{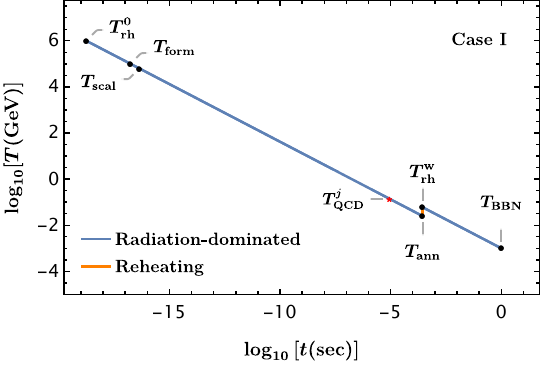} \hspace{3mm}
    \includegraphics[width=0.48\textwidth]{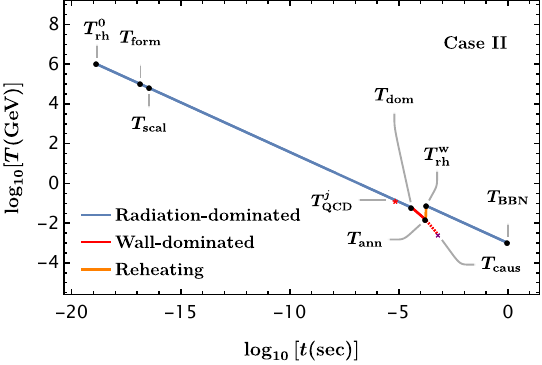}
    \caption{Schematic plots of early universe temperature evolution as a function of time. The left panel depicts purely radiation-dominated universe, while the right panel contains a short period of domain-wall dominance. The QCD phase transition time is labelled by the red star with $\TQCDj$ denoting a set of QCD phase transition temperatures of different $\theta$ domains. The purple cross in the right panel indicates the possible causal-limit temperature $\Tcaus$, which is not reached in this example. 
    }
    \label{fig:evolution}
\end{figure}

Before we discuss the domain wall evolution, we first list a few critical early-universe moments that are labeled by the corresponding temperatures.  
\bit
\item $\Trhi$: the reheating temperature immediately after inflation. 
\item $\Tform$: the initial domain wall formation temperature. Depending on the dynamics related to the order-parameter field $S$, a nontrivial first-order or second-order phase transition could happen at this time. It is around the discrete symmetry breaking scale or $\Tform \sim f$. 
\item $\Tscaling$: the time when domain walls reach the scaling behavior during a radiation-dominated universe. After this time, the domain walls have a fixed co-moving number density.  The domain wall curvature or the averaged separating distance $L$ scales linearly in time $t$. 
\item $\Tdomin$: the time when the domain walls dominate the total energy of the universe. Note that if the domain walls collapse before they dominate the universe, there is no domain-wall-dominated period. 
\item $\Tcaus$: the time when the wall separation speed reaches the causal limit or the speed of light. Any walls surviving after this time will not collapse.  
\item $\TQCDj$: the QCD phase transition temperature. Note that different domains labelled by $j$ have different $\theta_j$'s and hence different  $\TQCDj$'s as well as different orders of QCD phase transition. 
\item $\Tann$: the domain wall annihilation time. Due to the biased potential among different domains from QCD instanton effects, the vacuum energy drives the domain walls to collapse. After this time, only the $j=0$ or $\theta=0$ domain survives to the current universe. For $N>2$, a set of $\Tann$'s are anticipated because of different values of biased potential.
\item $\Trhw$: the ``wall reheating temperature" after the domain walls collapse and convert their energy to radiation. 
\eit
Depending on the relation between $\Tdomin$ and $\Tann$, two distinct situations can happen. When $\Tann > \Tdomin$ (denoted as Case I), the domain walls disappear before they dominate the universe. The whole domain wall evolution is approximately in the radiation-dominated universe and can be described by the scaling behavior till the annihilation time. The evolution of temperature as a function of time is plotted in the left panel of Fig.~\ref{fig:evolution}, where the rough scales for different temperatures are provided with $f \sim 10^5$~GeV.  For this example, the annihilation temperature is comfortably higher than the BBN temperature, so the BBN observables will not be affected by the early existence of domain walls. Also note that the wall reheating temperature is only slightly higher than $T_{\rm ann}$ because the energy contained in the walls is subdominant compared to the main radiation energy. 

For the other case with $\Tann < \Tdomin$ (denoted as Case II and illustrated in the right panel of Fig.~\ref{fig:evolution}), the temperature changes its dependence on time from the early $T \propto t^{-1/2}$ in the radiation-dominated universe to $T \propto t^{-2}$ after domain-wall dominance. After a period of a quick drop of temperature that reaches $\Tann$, the collapse of domain walls reheats the universe to $\Trhw$, which could be much higher than $\Tann$. For this case, $\Tann$ could be lower than $T_{\rm BBN}$, but with $\Trhw$ above $T_{\rm BBN}$ to guarantee a radiation-dominated universe before the BBN time.

\subsection{Details of domain wall evolution}
In this section, we detail the domain wall evolution through cosmic history. We consider $\Trhi > \Tform$, so that domain walls form and survive in the radiation-dominated universe after inflation and reheating. For $T<\Trhi$, we have the radiation energy density $\rho_{R}(T)=(\pi^{2}/30) \, g_{*}(T) T^{4}$ and the Hubble scale $H(T)=(\pi^{2}/90)^{1/2}\,g_{*}(T)^{1/2}\,T^{2}/\mpl $, where $g_{*}(T)$ denotes the total radiation degrees of freedom and $\mpl = 1/\sqrt{8 \pi G} = 2.43 \times 10^{18} \, \rm GeV$ is the reduced Planck mass with $G$ as the Newton constant. The scale factor during the radiation-dominated universe scales with time as $a \propto t^{1/2}$, and thus $H(t) = 1/(2t)$ and $T \propto t^{-1/2}$. We will remain agnostic regarding the order of the phase transition for the discrete symmetry breaking, as the information about the initial size of the domain wall population determined by the Kibble-Zurek mechanism~\cite{Kibble:1976sj,Zurek:1985qw} will be erased by the subsequent evolution. 

The evolution of domain walls is governed by two forces: the tension force $p_{T} = \rho_{w} = \sigma/L$, where $L$ is the curvature radius of the walls and controls the average distance between domain walls, and the friction force due to the reflection of plasma particles off domain walls. The wall network evolution, taking into account effects of the Hubble damping and the friction force, is determined by velocity-dependent one-scale (VOS) model~\cite{Martins:2016ois} which agrees with the numerical simulation. In this model the root mean square velocity of the wall $v$ and $L$ are related by the coupled differential equations as follows~\cite{Martins:2016ois}:
\beqa\label{eq:vos}
\dfrac{dL}{dt} &=& H L + v^{2}\dfrac{L}{l_{d}} + c_{w}v \,, \\ \vspace{3mm}
\dfrac{dv}{dt} &=& (1-v^{2})\Big{(}\dfrac{k_{w}}{L} - \dfrac{v}{l_{d}} \Big{)}\, ,
\eeqa
where $c_{w}$ and $k_{w}$ are constant phenomenological parameters determined by the numerical simulation and $l_{d}$ is the damping scale given by $l_{d}^{-1}= 3 H + l_{f}^{-1}$. Here $l_{f}$ is the friction length. If the friction force is negligible, which would be the case if the relativistic plasma particles perfectly transmit through the walls, then $l_{d}^{-1} = 3 H$ (see \cite{Blasi:2022ayo} for the case with dominant friction term and its effect on domain wall evolution). In such cases the scaling solution with $L=L_{0}t$ and $v= v_{0}$ is reached where $L_{0}$ and $v_{0}$ are constants determined by $c_{w}$ and $k_{w}$. For the case of a radiation-dominated universe, numerical simulations suggest that the scaling solution is reached at the temperature $\Tscaling\approx \Tform/30$, with $L_{0}=1.2$ and $v_{0}=0.42$ (which give $k_{w}=0.66$ and $c_{w}=0.81$)~\cite{Martins:2016ois}. While this result is only valid for $\mathbb{Z}_2$, for general $\mathbb{Z}_N$ the scaling solution is also seen in numerical simulations with $\rho_{w} = \mathcal{A} \,\sigma/t$ and $\mathcal{A}\approx 0.4 N$~\cite{Hiramatsu:2012sc, Kawasaki:2014sqa}. 

As the domain wall energy density in the scaling regime goes as $\rho_{w} \propto t^{-1}$, whereas the radiation energy density during the radiation-dominated era goes as $\rho_{R}\propto t^{-2}$, domain walls will eventually dominate the total energy density of the universe. This happens at the temperature $\Tdomin$  that is determined by $\rho_{w} = \rho_{R}$ and given by
\beqa
\Tdomin \approx 45 \, {\rm MeV} \, \Big{(}\frac{\mathcal{A}}{0.8}\Big{)}^{1/2}\,\left(\frac{\sigma}{10^{16}\,\mbox{GeV}^{3}}\right)^{1/2} \, \left(\frac{g_*(\Tdomin)}{10}\right)^{-1/4}\,.
\eeqa
Parametrically one has $t_{\rm dom} = 3 M_{\rm pl}^{2}/(4\mathcal{A}\sigma)$. 
As domain walls cannot dominate the energy density of universe till today or even BBN time, they would have to annihilate. There are two cases: I) walls annihilate before they dominate the total energy density ($\Tdomin < \Tann$); II) walls annihilate after they dominate the energy density of the universe ($\Tdomin > \Tann$). We will consider both cases below.

\subsubsection{Case I: $\Tdomin < \Tann$}
In this case the radiation energy dominates the universe until the walls annihilate. Thus, the scaling solution described above, with $\rho_{w}\propto t^{-1}$, is valid till the annihilation time. As different domains, with different $\theta_{j}$'s, have different potential energies as given by Eq.~\eqref{eq:potential}, this leads to a vacuum pressure force $p_{V} = V_{\rm bias}$, with $V_{\rm bias}$ being the vacuum energy difference between adjacent domains. Note that depending on the form of the effective potential, there could be a series of $V_{\rm bias}^{ij}$ labelled by two adjacent domain numbers. In this section, we keep $V_{\rm bias}$ as a general parameter and will come back to the indexed $V_{\rm bias}^{ij}$ later.

As $p_{T}$ tries to stretch the walls as depicted by the scaling solution, $p_{V}$ acts to collapse the domains with higher vacuum energies. The collapse starts when $p_{T} = p_{V}$, which corresponds to the temperature 
\begin{equation}\label{eq:Tann}
\Tann \approx 120 \, {\rm MeV} \, \left(\frac{V_{\rm bias}}{(100 \, \rm MeV)^{4}} \right)^{1/2}\, \left(\frac{\mathcal{A}}{0.8}\right)^{-1/2}\,\left(\frac{\sigma}{10^{16}\,{\rm GeV}^{3}}\right)^{-1/2} \, \left(\frac{g_{*}(\Tann)}{10}\right)^{-1/4}\, .
\end{equation} 
Parametrically one has $t_{\rm ann} = \mathcal{A}\, \sigma/V_{\rm bias}$. The larger the surface tension, the longer it takes to collapse the walls, while the larger $V_{\rm bias}$, the earlier annihilation happens. For a higher symmetry breaking scale or larger $\sigma$, it takes longer to annihilate the walls and thus leads to a lower $\Tann$. After annihilation the energy contained in the domain walls gets transferred to the radiation, reheating the universe to $\Trhw$ given by 
\begin{equation}
\Trhw = (1+\mathcal{F})^{1/4}\left(\frac{g_{*}(\Tann)}{g_{*}(\Trhw)}\right)^{1/4} \, \Tann \, ,
\end{equation}
where $\mathcal{F}$ is given by
\begin{equation}\label{eq:Ffactor}
\mathcal{F} = \frac{\rho_{w}(\Tann)}{\rho_{R}(\Tann)} \approx 0.14 \, \left(\frac{\mathcal{A}}{0.8}\right)^{2}\, \left(\frac{\sigma}{10^{16}\,\rm GeV^{3}}\right)^{2}\,\left(\frac{(100 \, \rm MeV)^{4}}{V_{\rm bias}} \right)\, .
\end{equation}
This implies that for most of the parameter space $\Tann \approx \Trhw$ for Case I. Given the tight constraints from BBN observables, we impose a constraint of $\Tann \approx \Trhw \gtrsim 3~\rm MeV$~\cite{Bai:2021ibt,Bringmann:2023opz}. Using Eq.~\eqref{eq:Tann}, this translates into an upper bound on $\sigma$ as
\beqa
\sigma < 1.6 \times 10^{19}\,\mbox{GeV}^3 \, \left(\frac{V_{\rm bias}}{(100 \, \rm MeV)^{4}} \right)\, \left(\frac{\mathcal{A}}{0.8}\right)^{-1}\, \left(\frac{g_{*}(\Tann)}{10}\right)^{-1/2} ~. 
\eeqa
Later we will show that this bound is practically irrelevant because domain-wall dominance and/or the causal-wall-separation limit will be reached before one reaches the above upper limit.


Note that we have used temperature-independent $V_{\rm bias}$ so far. 
In fact, for the potential in Eq.~\eqref{eq:potential} the thermal corrections are known for $T<\TQCDj$~\cite{Gasser:1986vb, Gasser:1987ah, GrillidiCortona:2015jxo} based on the chiral Lagrangian of the $\theta=0$ vacuum (for the GW spectra presented in Sec.~\ref{sec:GW}, we will check the finite-temperature effects). For $T>\TQCDj$ the $\theta$-dependent potential is not known reliably. The dilute instanton gas calculation~\cite{Gross:1980br}, relying on the finite-temperature perturbative QCD, is only valid for $T>10^{6}\,\rm GeV$~\cite{GrillidiCortona:2015jxo} [see \cite{Preskill:1991kd, Banerjee:2023hcx} for the domain wall annihilation assuming the validity of dilute instanton gas calculation up to $\mathcal{O}(\rm GeV)$]. In this paper we will consider the case with $\Tann < \TQCDj$ as we will be immune to the above uncertainties and it also opens up a new phenomenological study of QCD phase transition at finite QCD theta angle (see Sec.~\ref{sec:phase-transition} for details). For the case with $\Tann>\TQCDj$, we would have $\theta=0$ in all Hubble patches at $T_{\rm QCD}$  and the QCD phase transition would be a ``boring" crossover one at $T_{\rm QCD}\approx 170~\rm MeV$~\cite{Fodor:2001pe}. If we demand that $\Tann < \TQCDj \approx 125 \, \rm MeV$ (see Fig.~\ref{fig:phase:QCD}), we obtain a lower bound on $\sigma$
\begin{equation}
\label{eq:siglowQCD}
\sigma \geq 9.2 \times 10^{15} \, {\rm GeV}^{3} \, \left(\frac{\mathcal{A}}{0.8}\right)^{-1}\, \left(\frac{V_{\rm bias}}{(100 \, {\rm MeV})^{4}} \right)\, \left(\frac{g_{*}(\Tann)}{10}\right)^{-1/2}\,.
\end{equation}

Finally, we demand $\Tdomin < \Tann$ for this case so that domain walls annihilate under radiation domination. This imposes an upper bound on $\sigma$ given by
\begin{equation}\label{eq:sigupRD}
\sigma \leq 2.6 \times 10^{16} \, {\rm GeV}^{3} \, \left(\frac{\mathcal{A}}{0.8}\right)^{-1}\, \left(\frac{V_{\rm bias}}{(100 \, \rm MeV)^{4}} \right)^{1/2}\,.
\end{equation}
For $\sigma$ larger than the above value, the universe will enter the domain-wall-dominated era before wall annihilation. 

Comparing the two constraints from opposite ends in Eq.~\eqref{eq:siglowQCD} and Eq.~\eqref{eq:sigupRD}, one can see that for Case I there is only a small parameter space in $\sigma$ that allows for potential non-trivial QCD phase transitions in different domains unless a smaller value of $V_{\rm bias}$ is given.

\subsubsection{Case II: $\Tdomin > \Tann$}

In this case when $t>t_{\rm dom}$, the scaling solution given in the previous subsection with $L\propto t$ is no longer valid. In the domain-wall-dominated universe, one has the equation of state $\omega=-2/3$~\cite{Friedland:2002qs} and $H = \sqrt{\rho_{w}/(3\mpl^{2})}=\sqrt{\sigma/(3 L \mpl^{2})}$. Substituting it into Eq.~\eqref{eq:vos}, with the boundary conditions  $L(t_{\rm dom}) \equiv L_{\rm dom} = t_{\rm dom}/\mathcal{A}$ and $v\simeq 0$, 
\begin{equation}\label{eq:LtWD}
L(t) = \left(\dfrac{t-t_{\rm dom}}{2 \sqrt{L_{\rm caus}}} + \sqrt{L_{\rm dom}} \right)^{2} \, ,
\end{equation}
where we have defined the causal wall separation distance $L_{\rm caus} = 3 \mpl^{2}/\sigma$, which is the average domain wall distance at the causality limit $L = H^{-1} = L_{\rm caus}$. The time at which $L \rightarrow L_{\rm caus}$ is  given by $t_{\rm caus} = t_{\rm dom} (8\mathcal{A}-3)$. For $t>t_{\rm caus}$, we have $L>H^{-1}$, \ie, domain walls are separated by a scale larger than the horizon size. In such a case the assumptions of Friedmann–Robertson–Walker (FRW) cosmology, homogeneity and isotropy, break down. Moreover, the vacuum energy inside the domain would likely determine the Hubble evolution and guide the universe into a period of inflation. Accordingly, the domain walls cannot collapse, and thus this scenario with $L>H^{-1}$ is clearly ruled out. Hence, we demand that domain walls annihilate before the causality limit is reached, \ie, $t_{\rm ann} < t_{\rm caus}$.  

In the domain-wall-dominated scenario, we can derive from the annihilation condition $\rho_{w} = V_{\rm bias}$
\begin{equation}
t_{\rm ann} = t_{\rm dom} + 2 \sqrt{L_{\rm caus}}\left[\left(\frac{\sigma}{V_{\rm bias}}\right)^{1/2} - L_{\rm dom}^{1/2} \right] \,.
\end{equation}
Requiring $t_{\rm ann} < t_{\rm caus}$ then imposes an upper bound on $\sigma$
\begin{equation}\label{eq:sigupwd}
\sigma < 4.2 \times 10^{16} \, \rm GeV^{3} \, \left(\frac{V_{\rm bias}}{(100 \, \rm MeV)^{4}} \right)^{1/2} \, .
\end{equation}
Combined with Eq.~\eqref{eq:sigupRD}, the domain-wall-dominated universe happens for $\sigma$ satisfying
\begin{equation}
\label{eq:domain-wall-dominance}
 2.6 \times 10^{16} \, {\rm GeV}^{3} \, \left(\frac{\mathcal{A}}{0.8}\right)^{-1}\, \left(\frac{V_{\rm bias}}{(100 \, \rm MeV)^{4}} \right)^{1/2} < \sigma < 4.2 \times 10^{16} \, {\rm GeV}^{3} \, \left(\frac{V_{\rm bias}}{(100 \, \rm MeV)^{4}} \right)^{1/2} \, .
\end{equation}

After domain walls annihilate, the energy contained in the wall network is transferred to radiation and reheats the universe to a temperature $\Trhw$ of 
\begin{equation}\label{eq:TwrhWD}
\Trhw \approx 74 \,{\rm MeV} \, \left(\frac{g_{*}(\Trhw)}{10}\right)^{-1/4} \, \left(\frac{V_{\rm bias}}{(100 \, \rm MeV)^{4}} \right)^{1/4} \,,
\end{equation}
which is obviously higher than the BBN temperature of $\mathcal{O}(1\,\mbox{MeV})$ and thus safe from the BBN constraints.

To derive the values of other characteristic temperatures, we use  Eq.~\eqref{eq:LtWD} and $H = \dot{a}/a = 2/(t+3\,t_{\rm dom})$, which give
\begin{equation}
a(t) = a(t_{\rm dom}) \left(\frac{t}{4\,t_{\rm dom}} + \frac{3}{4} \right)^{2} \, , \qquad T(t) = \Tdomin \left(\frac{t}{4\,t_{\rm dom}} + \frac{3}{4}\right)^{-2} \, .
\end{equation}
At $t=t_{\rm ann}$, the domain wall annihilation temperature is given by 
\begin{equation}
\Tann = 60 \, {\rm MeV} \, \left(\frac{\mathcal{A}}{0.8}\right)^{-3/2} \, \left(\frac{g_{*}(\Tdomin)}{10}\right)^{-1/4} \, \left(\frac{V_{\rm bias}}{(100 \, \rm MeV)^{4}} \right) \, \left(\frac{\sigma}{3 \times 10^{16}\,\rm GeV^{3}}\right)^{-3/2} \,.
\end{equation}
Together with the condition given in Eq.~\eqref{eq:domain-wall-dominance}, one can see that $\Tann < \TQCDj$ and therefore non-trivial QCD phase transitions could happen in Case II. 

At $t = t_{\rm caus}$, one has
\begin{equation}
\Tcaus = \frac{\Tdomin}{4\,\mathcal{A}^{2}} = 30\, {\rm MeV} \, \left(\frac{\sigma}{3 \times 10^{16} \, \rm GeV^{3}}\right)^{1/2} \, \left(\frac{\mathcal{A}}{0.8}\right)^{-3/2} \, \left(\frac{g_{*}(\Tdomin)}{10}\right)^{-1/4} \,.
\end{equation}  
Demanding $\Tann>\Tcaus$ gives the same upper bound on $\sigma$ as in Eq.~\eqref{eq:sigupwd}. Note that one can have $\Tcaus$ or $\Tann$ less than $T_{\rm BBN}$, since the universe that is reheated after wall annihilation has $\Trhw>T_{\rm BBN}$. The lowest value of $\Tcaus$ can be calculated by taking the smallest possible value of $\sigma$ [from Eq.~\eqref{eq:sigupRD}] that would cause domain wall domination, which is 
\begin{equation}
T_{\rm caus}^{\rm low} \approx 26 \, {\rm MeV} \, \left(\frac{\mathcal{A}}{0.8}\right)^{-2} \, \left(\frac{g_{*}(\Tdomin)}{10}\right)^{-1/4} \, \left(\frac{V_{\rm bias}}{(100 \, {\rm MeV})^{4}} \right)^{1/4} \,.
\end{equation}
For a large $\mathcal{A}$, \ie, for a large $N$ of the $\mathbb{Z}_{N}$ discrete symmetry, one can have a much smaller $T_{\rm caus}^{\rm low}$. 

\begin{figure}[th!]
\centering
    \includegraphics[width=0.48\textwidth]{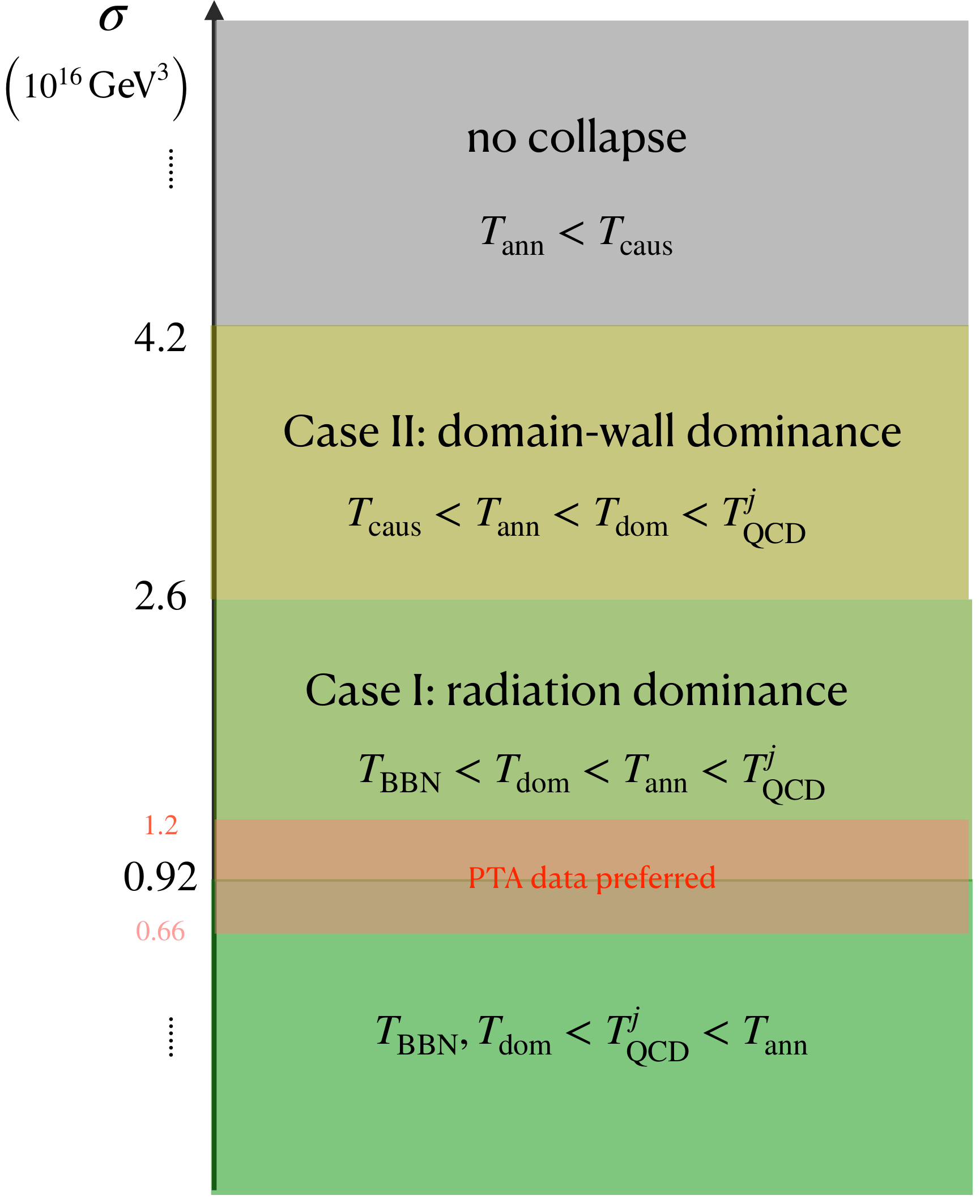}
    \caption{Different situations for different values of $\sigma$ with $\mathcal{A}=0.8$ and $V_{\rm bias} = (100\,\mbox{MeV})^4$. The PTA GW results prefer a value of $\sigma$ that is in the radiation dominance case with the annihilation temperature comparable to the QCD phase transition temperature (see Sec.~\ref{sec:pulsar-data} for details). Here, $V_{\rm bias}^{\rm max}$ is assumed to be valid up to $\Tann=148\,\rm{MeV}$.
    }
    \label{fig:sigmaschem}
\end{figure}

We summarize the different situations for different values of $\sigma$ in Fig.~\ref{fig:sigmaschem}. In this figure, we also show the preferred range of $\sigma$ by the PTA data, which turns out to sit near the boundary between radiation dominance and wall annihilation before QCD phase transition. 

\section{QCD phase transition with non-zero $\theta$}
\label{sec:phase-transition}

The QCD with $\theta=0$ in the SM has been shown to have a crossover when it transits from the high-temperature quark-gluon plasma phase to the low-temperature hadronic phase~\cite{Aoki:2006we,Bhattacharya:2014ara}. For a non-zero $\theta$, there is no robust calculation for the strength of finite-temperature phase transitions. On the other hand, it has been pointed out a long time ago by Dashen~\cite{Dashen:1970et} and later by Witten~\cite{Witten:1980sp} using Large-$N_c$ expansion that CP is spontaneously broken for $\theta=\pi$ with a first-order transition along the $\theta$ direction. More recently, this result is obtained and confirmed by using the generalized symmetry tool and 't Hooft anomaly matching~\cite{Gaiotto:2017yup,Gaiotto:2017tne}. This result also suggests QCD works quite differently for $\theta = \pi$  or close to $\pi$.   

Without a non-perturbative tool to analyze the QCD phase diagram with a non-zero $\theta$, one could use some phenomenological model to gain some insights about the QCD finite-temperature phase transition. In Appendix~\ref{sec:appendix}, we adopt the so-called linear sigma model coupled to quarks (LSM$q$) introduced in Ref.~\cite{Pisarski:1996ne}. For two quark flavors, the finite-temperature phase transition with a nonzero $\theta$ has been studied in Ref.~\cite{Mizher:2008hf}. A comparison between the LSM$q$ and Nambu-Jona-Lasinio (NJL) model has been studied in Ref.~\cite{Boomsma:2009eh}. In Appendix~\ref{sec:appendix}, we perform a detailed study about finite-temperature phase transitions for the LSM$q$ model with three quark flavors and matched meson spectra. 

\begin{figure}[th!]
\centering
    \includegraphics[width=0.65\textwidth]{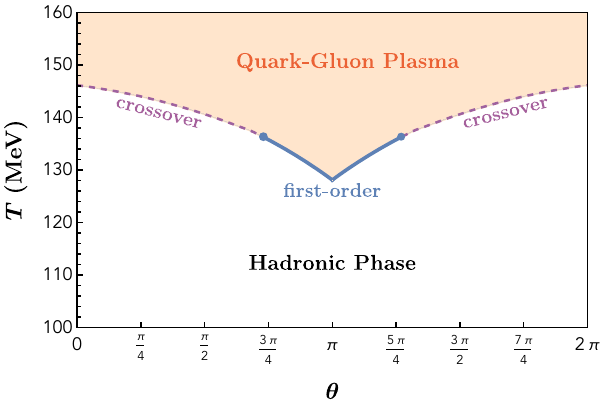}
    \caption{The phase diagram in $\theta$ and $T$ based on the phenomenological 3-flavor LSM$q$ model (see Appendix~\ref{sec:appendix} for detailed calculations).
    The critical points are located at $(\theta, T) = (0.7\pi, 136\,\mbox{MeV})$ and $(1.3\pi, 136\,\mbox{MeV})$.
    }
    \label{fig:phase:QCD}
\end{figure}




The results in Appendix~\ref{sec:appendix} are summarized in the phase diagram of $\theta-T$ in Fig.~\ref{fig:phase:QCD}, where we show the phase transition temperatures for different $\theta$'s. For $\theta = 0$, the crossover transition is recovered. The phase transition temperature is around 146~MeV, which is close to the Lattice QCD result ($\approx 170$~MeV)~\cite{Fodor:2001pe}. For $\theta$ close to $\pi$, first-order phase transitions happen with a lower phase-transition temperature as $\theta$ gets closer to $\pi$. The critical points (labelled by the blue dots) are located at $(\theta, T) = (0.7\pi, 136\,\mbox{MeV})$ and $(1.3\pi, 136\,\mbox{MeV})$. Therefore, for $\theta\in[\theta_c,2\pi-\theta_c]$ with $\theta_c\approx 0.7\pi$, the QCD phase transition is first-order based on the LSM$q$ model. We want to emphasize again that the LSM$q$ model is just a phenomenological model to provide the hint of a first-order QCD phase transition. The true phase diagram could be similar to the one in Fig.~\ref{fig:phase:QCD}, although the precise locations of critical points or the value of $\theta_c$ are subject to additional theoretical and numerical studies. 

If $\Tann<\TQCDj$, different domains with different $\theta_j$'s could undergo phase transitions before the domain walls are annihilated due to the biased potential. For those domains with $\theta_j\in[\theta_c,2\pi-\theta_c]$, the QCD phase transitions are first-order and could generate additional GW's beyond the one generated by domain wall collapse. Later, we will show that the PTA observed GW prefers a range of $\sigma$ with $\Tann<\TQCDj$, or a non-trivial QCD phase transition.

\section{Gravitational wave signatures}
\label{sec:GW}
In this section we will consider GW emissions from three different sources 1) collapse of the domain wall network, 2) QCD phase transition, and 3) phase transition from the discrete symmetry breaking at $\Tform$. While the generation of GW's from the last two sources depend on the model parameters and could be absent, GW emission from domain wall annihilation is guaranteed if the PTA observed GW is due to domain-wall annihilation.
As we will show, the GW's from above cases will span more than ten orders of magnitude in frequency, and could be probed with future GW experiments. 

\subsection{GW from domain wall collapse}
As we pointed out in Sec.~\ref{sec:evolution}, after formation domain walls evolve to the scaling regime and annihilate due to the vacuum energy difference across the wall; the walls could annihilate under radiation domination $\Tann>\Tdomin$, or during the domain-wall-dominated era $\Tann<\Tdomin$. In this paper we will mainly focus on the former case but will also estimate the frequency and amplitude of emitted GW's in the latter case. 

The estimation of GW amplitude could be performed using the Einstein's quadrupole formula~\cite{Einstein:1918btx} with the gravitational radiation power as $P_{\rm {GW}}\sim G\, (d^{3}Q/dt^{3})^{2}$. Here, $Q$ is the transverse-traceless part of the quadrupole moment of matter. Two time-derivatives of this formula come from using the tensor virial theorem to convert the volume integration of the spatial part of energy-momentum tensor into double time-derivatives of the quadrupole moment. The third time-derivative is simply from calculating the propagating graviton field. For the domain wall case~\cite{Gleiser:1998na}, one has $Q\sim M_{\rm {wall}} L(t)^{2}$ with $L$ being the curvature radius of the wall and the mass of the wall $M_{\rm wall} \sim \sigma L^{2}$. The energy density of GW's released per unit Hubble time $\rho_{\rm GW} \sim  P_{\rm GW}H^{-1}/L^{3}$, where the factor $L^{-3}$ can be thought of as  the number density of GW sources. We will evaluate $\rho_{\rm GW}$ separately for the radiation domination and the domain-wall domination cases. The peak-frequency of the GW at the annihilation time, for both cases, is given by $f(t_{\rm ann}) \sim H(t_{\rm ann})$.

\subsubsection{Case I: collapse during radiation-dominated era}

In the case of domain wall annihilation during the radiation-dominated era, we can use $L(t) \propto t$ to obtain $P_{\rm GW}\propto G \,\sigma^{2} \,t^{2}$, which leads to $\rho_{\rm GW}\sim G \,\sigma^{2}$. The overall constant can be fixed by comparing to the spectrum obtained through numerical simulations~\cite{Gleiser:1998na,Hiramatsu:2012sc,Hiramatsu:2013qaa} (see Ref.~\cite{Saikawa:2017hiv} for a review). In particular, the spectrum of GW's per unit logarithmic frequency interval has a peak value 
\begin{equation}\label{eq:rhogwDW}
    \dfrac{d\rho_{\rm GW}}{d\ln k}\Big{|}_{\rm peak} = \tilde{\epsilon}_{\rm GW}G \mathcal{A}^{2}\sigma^{2} \, .
\end{equation}
Here, $k$ is the comoving wave number. 
The peak amplitude at $t_{\rm ann}$ is then given by
\begin{equation}\label{eq:gwampann}
    \Omega_{\rm GW}(t_{\rm ann}) = \frac{1}{\rho_{c}(t_{\rm ann})} \left( \frac{d\rho_{\rm GW}(t_{\rm ann})}{d\ln k} \right)\Big{|}_{\rm peak} = \frac{8\pi \tilde{\epsilon}_{\rm GW} G^{2} \mathcal{A}^{2} \sigma^{2}}{3 H(t_{\rm ann})^{2}} \, ,
\end{equation}
with $\rho_{c}(t_{\rm ann})$ as the critical energy density at $t_{\rm ann}$. Here, $\tilde{\epsilon}_{\rm GW}=0.7 \pm 0.4$ is a phenomenological parameter determined by numerical simulations~\cite{Hiramatsu:2013qaa}. The peak frequency of the spectrum has $f_{\rm peak}(t_{\rm ann}) \approx H(t_{\rm ann})$ based on the simulation up to $N = 6$~\cite{Hiramatsu:2013qaa}.

After GW productions at the domain wall annihilation time, the amplitude and frequency of the GW are red-shifted till today. The peak amplitude today ($t_{0}$) is given by~\cite{Saikawa:2017hiv}
\begin{align}\label{eq:omegagwDW}
 \Omega_{\rm GW}h^{2}(t_{0})\Big{|}_{\rm peak} =  3 \times 10^{-8} \, \left(\frac{\tilde{\epsilon}_{\rm GW}}{0.7} \right)\, \left(\frac{\mathcal{A}}{0.8}\right)^{2} \, \left(\frac{\sigma}{10^{16} \, \rm GeV^{3}}\right)^{2} \, \left(\frac{\Tann}{100 \, \rm MeV}\right)^{-4}\, \left(\frac{g_{*s}(\Tann)}{10}\right)^{-4/3}\,,
\end{align}
and the peak frequency is given by
\begin{equation}\label{eq:fgwDW}
    f_{\rm peak} = 1.1 \times 10^{-8} \, {\rm Hz} \, \left(\frac{g_{*}(\Tann)}{10}\right)^{1/2}\, \left(\frac{g_{*s}(\Tann)}{10}\right)^{-1/3}\, \left(\frac{\Tann}{100 \, \rm MeV}\right)\, ,
\end{equation}
where $g_{*s}\approx g_{*}$ stands for the effective relativistic degrees of freedom for the entropy density.  The numerical simulations also find that the GW spectrum scales as $f^{3}$ for $f<f_{\rm peak}$, as expected from the causality argument~\cite{Caprini:2018mtu,Cai:2019cdl}, and as $f^{-1}$ for $f>f_{\rm peak}$~\cite{Hiramatsu:2013qaa} (the numerical simulation also shows a harder spectrum for a larger $N$~\cite{Hiramatsu:2012sc}), which is also quoted in Ref.~\cite{NANOGrav:2023hvm} as a case of the parametrization $S(x)=(a+b)^c/(bx^{-a/c}+ax^{b/c})^c$ with $a=3$ and $b\simeq c\simeq 1$.

Note that we are using $\Tann$ and $\sigma$ as model parameters to describe the GW amplitude and frequency. While $\Tann$ is determined using $V_{\rm bias}$ and $\sigma$ (see Eq.~\eqref{eq:Tann}), $V_{\rm bias}(T)$ could be a function of temperature. In fact, the finite temperature corrections to the potential in Eq.~\eqref{eq:thetapotential} are evaluated for $T<T_{\rm QCD}$ and are given by~\cite{GrillidiCortona:2015jxo}
\begin{equation}
   \frac{V\Big(\langle S\rangle_j; T\Big)}{V\Big(\langle S\rangle_j\Big)} = 1 + \frac{3}{2}\frac{T^{4}}{f_{\pi}^{2}\,m_{\pi}^{2}\,(\langle S\rangle_j)}J_{0}\left[\frac{m_{\pi}^{2}(\langle S\rangle_j)}{T^{2}} \right]\, ,
\end{equation}
where $J_{0}(x)=-\frac{1}{\pi^{2}}\int_{0}^{\infty}\,dq\,q^{2}\,\log\big{(}1 - e^{-\sqrt{q^{2}+x}} \big{)}$ and $m_{\pi}^{2}(\langle S\rangle_j)  = - V\Big(\langle S\rangle_j\Big)/f_\pi^2$
with $m_{\pi}=135\,\rm MeV$ being the pion mass in the normal ($j=0$) vacuum. The finite temperature correction changes $V_{\rm bias}$ by around $10\%$ for $T\approx 125\,\rm MeV$, above which we expect that the QCD phase transition could change the potential drastically~\cite{GrillidiCortona:2015jxo}. Using Eq.~\eqref{eq:Tann}, we expect a max $5\%$ change in $\Tann$. This will lead to a maximum of $5\%$ and $20\%$ change in the peak frequency and amplitude, respectively. Given the larger uncertainty in other parameters ($\tilde{\epsilon}_{\rm GW}$ and $\mathcal{A}$), we will neglect the uncertainty from finite temperature corrections. As a result, we will use the $T=0$ potential in Eq.~\eqref{eq:thetapotential} to calculate $V_{\rm bias}$. 

Note that while the spectral characteristics of GW from domain wall annihilation mentioned above were obtained for the $\mathbb{Z}_{2}$ case, for $\mathbb{Z}_{N}$ the results should be obtained with some slight modifications. For the general $\mathbb{Z}_{N}$ case, the energy difference $V_{\rm bias}^{ij}$ between two adjacent domains (labelled by $\theta_{i}$ and $\theta_{j}$) could be the same or different. Similar situations are also pointed out in \cite{Higaki:2016jjh} and \cite{Wu:2022stu} under the context of $U(1)_{\rm PQ}^N$ and $\mathbb{Z}_N$ symmetries, respectively, in the latter of which the $\mathbb{Z}_3$ case is studied in detail. If $V_{\rm bias}^{ij}$ are all the same, all domain walls collapse at around the same time with a common $\Tann$. Then, there is a single peak in gravitational wave spectrum. In this case, according to numerical simulations~\cite{Hiramatsu:2012sc}, the peak amplitude and frequency match with $\mathbb{Z}_{2}$. While the spectrum at low $f<f_{\rm peak}$ is same as the $\mathbb{Z}_{2}$ case, for $f>f_{\rm peak}$ the amplitude is slightly enhanced relative to the $\mathbb{Z}_{2}$ case~\cite{Hiramatsu:2012sc}. This enhancement for a large $N$ at higher frequencies is caused by the production of many configurations with sub-Hubble sizes, which after collapse lead to higher frequency GW's~\cite{Saikawa:2017hiv}.   

For the case with various $V_{\rm bias}^{ij}$ values, which could be the case for non-linear potentials such as the one in Eq.~\eqref{eq:thetapotential}, one has sequential annihilation of domain walls: domain walls with a larger $V_{\rm bias}^{ij}$ collapse earlier with a higher $\Tann$. 
Since different $\Tann$'s lead to different GW  peak frequencies, a multi-peak GW spectrum is anticipated from the sequential annihilation of walls. The amplitude of GW produced at some $\Tann$ is determined by the fraction of the total number of domain walls undergoing collapse at that temperature. This fraction is determined by the statistical distribution of $V_{\rm bias}^{ij}$ for a given discrete symmetry $\mathbb{Z}_{N}$. 

\begin{figure}[th!]
\centering
    \includegraphics[width=0.48\textwidth]{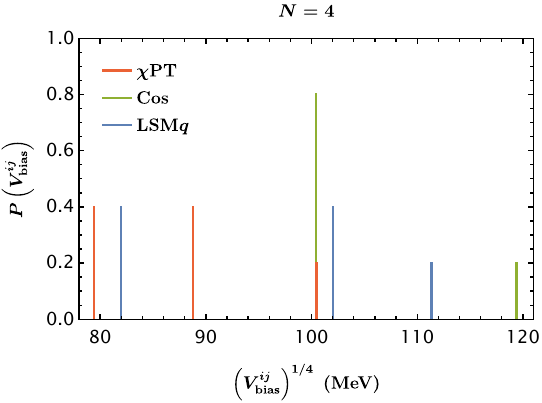}\hspace{3mm}
    \includegraphics[width=0.48\textwidth]{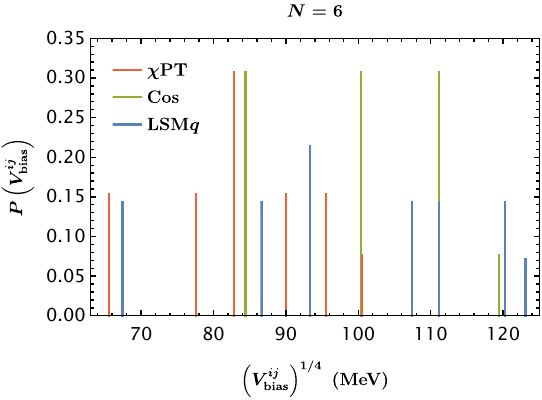}
    \caption{Statistical distributions of $(V_{\rm bias}^{ij})^{1/4}$ for $N=4$ (left) and $N=6$ (right) for the ``$\chi$PT'' (red), ``Cos'' (green), and LSM$q$ (blue) potentials.
    }
    \label{fig:Vij:dist}
\end{figure}

The situation for $\mathbb{Z}_2$ is very simple with $V_{\rm bias} = (100.4\,\mbox{MeV})^4$. For  $\mathbb{Z}_{4}$, one has domains with $\theta_j={0,\pi/2,\pi,3\pi/2}$ which we denote by $j=0,1,2,3$. Using the potential in Eq.~\eqref{eq:thetapotential}, one has $V_{\rm bias}^{01}=V_{\rm bias}^{03}\approx (79.4\,\rm MeV)^{4}$, $V_{\rm bias}^{12}=V_{\rm bias}^{23}\approx (88.8 \,\rm MeV)^{4}$, and $V_{\rm bias}^{02}\approx (100.4 \,\rm MeV)^{4}$, where $V_{\rm bias}^{ij}=|V(\theta_{i})-V(\theta_{j})|$. As initial domains, \ie, $j=0,1,2,3$, are created with equal probabilities, the statistical distribution of $V_{\rm bias}^{ij}$ for the $N=4$ case can be obtained, as shown in the left panel of Fig.~\ref{fig:Vij:dist} and labelled by ``$\chi$PT". For comparison, we also show the probability distributions using the effective potential constructed from the $\cos$ function: $V(\langle S_{j} \rangle) \approx V_{0}\, (1 - \cos(2\pi j/N))$ with $V_{0}=V_{\rm bias}^{\rm max}\approx (100.4 \, \rm MeV)^{4}$, and the one in the LSM$q$ model. In the right panel of Fig.~\ref{fig:Vij:dist}, we show the probability distributions for the $N=6$ case. For this case, there are potentially 15 $V_{\rm bias}^{ij}$'s in total. Two of them, $V_{\rm bias}^{15}$ and $V_{\rm bias}^{24}$, are zero and thus not included in the right panel  of Fig.~\ref{fig:Vij:dist}. Depending on the surrounding wall evolution, those domain walls have a more complicated collapsing possibility and are not included in our later analysis. A similar treatment is also used for the $N=4$ case by not including the $V^{13}_{\rm bias}$ wall in the later analysis.   

\begin{figure}[th!]
\centering
    \includegraphics[width=0.48\textwidth]{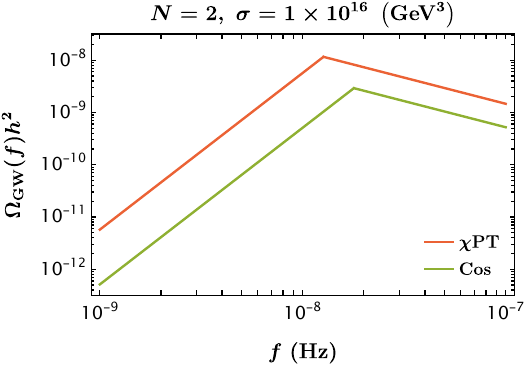} \hspace{3mm}
    \includegraphics[width=0.48\textwidth]{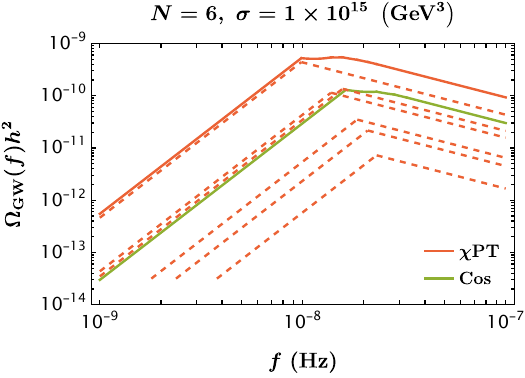}
    \caption{The GW spectra induced by the (sequential) domain-wall collapses based on the ``$\chi$PT'' (red) and ``Cos'' (green) potentials for $N=2$ (left) and $N=6$ (right). The dashed lines in the right panel denote the spectra induced by the collapses of individual walls based on the ``$\chi$PT'' potential, and the red solid line denotes the total spectrum after taking into account the statistical distribution. 
    }
    \label{fig:GW:Ann:XPT}
\end{figure}

After knowing the probability distributions of $V_{\rm bias}^{ij}$, we can then estimate the GW spectrum. For simplicity, we just sum the GW's generated from collapsed domain walls at different temperatures according to the specific $V_{\rm bias}^{ij}$ and Eq.~\eqref{eq:Tann} with $\mathcal{A}=1.6$ for $N=4$. With this assumption, we ignore the potential additional wall evolution after some fraction of walls are collapsed. Using the $\chi$PT and $N=6$ model as an example and with $\sigma=1\times10^{15}~{\rm GeV}^3$, we have $1/13$ of walls collapse at around $\Tann\approx 187$~MeV, $2/13$ of walls collapse at $\Tann\approx 169$~MeV, $150$~MeV, $112$~MeV, $79.6$~MeV, and $4/13$ of walls collapse at $\Tann\approx 127$~MeV. For each fraction, the GW amplitude is calculated using Eq.~\eqref{eq:gwampann} and the specific $\Tann$ including the spectrum feature of $f^{3}$ for $f<f_{\rm peak}$ and as $f^{-1}$ for $f>f_{\rm peak}$. Following this treatment, we show both the total and sub-components of GW spectra in Fig.~\ref{fig:GW:Ann:XPT} for $N=2$ (left panel) and $N=6$ (right panel) and for both $\chi$PT and $\mbox{Cos}$ potentials for comparison. For the $N=6$ case, one can see that the GW spectrum is dominated by the smallest $V_{\rm bias}^{ij}$ with the smallest $\Tann$ due to the larger inverse power dependence of the GW amplitude in $\Tann$ [see Eq.~\eqref{eq:omegagwDW}]. Furthermore, this spectrum also has an interesting plateau-like feature that could be used to identify the specific discrete symmetry group with a more precise measurement of the GW spectrum. 

\subsubsection{Case II: collapse during domain-wall-dominated era}
The GW spectrum analysis till now only considered the wall annihilation during the radiation-dominated period. Now, we turn to the GW generation from wall collapses during the wall-dominated period. Using $L(t)$ from Eq.~\eqref{eq:LtWD}, one has 
\begin{equation}
    P_{\rm GW} = \frac{32 \kappa}{3\pi\Mpl^{8}}\sigma^{5}L^{5} \, ,
\end{equation}
where $\kappa$ is a constant. The energy density of GW's released per unit Hubble time is 
\begin{equation}
    \rho_{\rm GW} = \frac{32\sqrt{3}\kappa }{3 \mpl^{7}}\sigma^{9/2}L^{5/2} \, .
\end{equation}
If the GW was emitted at $t=t_{\rm dom}$, then we can compare the above equation to the corresponding one from the radiation-dominated regime in Eq.~\eqref{eq:rhogwDW} to fix $\kappa=\frac{\pi}{9} \tilde{\epsilon}_{\rm GW}\mathcal{A}^{7}$. Similar to the case of radiation domination, the peak frequency of GW is at $f_{\rm peak} = H(t_{\rm ann}) = H(\Trhw)$ with an amplitude of
\begin{equation}
    \Omega_{\rm GW}(t_{\rm ann}) = \Omega_{\rm GW}(t_{\rm rh}^{\rm w}) = \frac{32\sqrt{3}}{81}\frac{\tilde{\epsilon}_{\rm GW}\mathcal{A}^{7}\sigma^{7}}{\mpl^{9}V_{\rm bias}^{5/2}H(\Trhw)^{2}}\, ,
\end{equation}
where we have used the fact that for instantaneous reheating $H(t_{\rm ann})=H(\Trhw)$ with $\Trhw$ given by Eq.~\eqref{eq:TwrhWD}. Taking into account the redshift factor from $\Trhw$ to today, the peak amplitude is
\begin{equation}
    \Omega_{\rm GW}h^{2}(t_{0})\Big{|}_{\rm peak}  \approx 6 \times 10^{-7} \, \, \Big{(}\frac{\tilde{\epsilon}_{\rm GW}}{0.7} \Big{)}\, \Big{(}\frac{\mathcal{A}}{0.8}\Big{)}^{7} \, \Big{(}\frac{\sigma}{3 \times 10^{16} \, \rm GeV^{3}}\Big{)}^{7} \, \Big{(}\frac{\Trhw}{100 \, \rm MeV}\Big{)}^{-14}\, \Big{(}\frac{g_{*}(\Trhw)}{10}\Big{)}^{-23/6}\,,
\end{equation}
and the peak frequency is 
\begin{equation}
    f_{\rm peak} = 1.1 \times 10^{-8} \, {\rm Hz} \, \left(\frac{g_{*}(\Trhw)}{10}\right)^{1/6}\,  \left(\frac{\Trhw}{100 \, \rm MeV}\right) \,.
\end{equation}
In above equations we have taken $g_{*s}(\Trhw)= g_{*}(\Trhw)$. 

\subsection{GW from QCD phase transition}
As mentioned in Sec.~\ref{sec:phase-transition}, the LSM$q$ model suggests a possibility of QCD first-order phase transition (PT) in the regions with non-zero $\theta$ values, in particular for  $\theta\in  [\theta_c, 2\pi - \theta_c]$ with $\theta_{c}\approx 0.7 \pi$ (see Fig.~\ref{fig:phase:QCD}) with the phase transition temperature of $\approx 125 \, \rm MeV$. If $\Tann \lesssim 125 \, \rm MeV$, then we expect to have QCD first-order PT for the domains with $\theta\in  [\theta_c, 2\pi - \theta_c]$ and GW productions.
As we do not have a trustworthy model to describe the QCD phase transition dynamics, we will follow a model-independent approach to describe the GW spectrum with the nucleation temperature fixed at $\approx 125 \, \rm MeV$. Fig.~\ref{fig:phase:QCD} suggests that only in the domains with $\theta\in  [\theta_c, 2\pi - \theta_c]$ are the FOPT and thus generation of GW's enabled. For a general $\mathbb{Z}_{N}$ case, as we expect to have domains with $\theta=2\pi j/N$ with $0\leq j\leq N-1$, the domains with $\lfloor N - N\theta_{c}/2\pi \rfloor \geq j \geq \lceil N\theta_{c}/2\pi \rceil $ will undergo QCD FOPT. Since at the time of phase transition we expect each $\theta_j$ to cover $1/N$ fraction of the universe, we expect FOPT in only a fraction of universe, denoted by $\zeta$, populated by domains satisfying the above condition on $j$. For $\theta_{c}\approx 0.7 \pi$, one has $\zeta=1/2,0,1/4,2/5,1/6$ for $N=2,3,4,5,6$, respectively. Those fractions will be the multiplication factors for the standard GW amplitudes from FOPT, where one has the whole universe undergoing the phase transition. 

The GW's from FOPT are mainly described by two parameters: $\alpha_{\rm GW}\approx \Delta V(T_{n})/\rho_{R}(T_{n})$ denotes the strength of the PT, and $\beta_{\rm GW}$ denotes the rate of bubble nucleation with $\beta_{\rm GW}/H(T_{n})$ being the commonly used parameter. Here $\Delta V(T_{n})$ denotes the vacuum energy difference between the false and true vacua at the nucleation temperature $T_{n}$. The SGWB from FOPT can come from three different processes: collision of bubbles, sound waves generated from bubble expansion, and turbulence in plasma. The GW spectra from these three processes have been evaluated numerically (see Appendix~\ref{sec:PTfomula} for formulas). Note that the $\zeta$ factor will act as a multiplicative suppression factor for the standard formulas from all three sources (see \cite{Caprini:2018mtu} for a recent review), as we have only $\zeta$ fraction of the universe undergoing FOPT.
Since determining the precise values of $\alpha_{\rm GW}$ and $\beta_{\rm GW}$ for QCD FOPT with a non-zero $\theta$ is challenging, we take model-independent approach and show the spectra for a range of parameters. In Fig.~\ref{fig:GW:spectroscopy} we show the spectra for $(\alpha_{\rm GW}, \beta_{\rm GW}/H(T^{\rm QCD}_n)) = (0.5,10^{4}), (0.5,10^{5})$ with $\zeta=0.5$ and compare them with the sensitivity curves of GW experiments.

\subsection{GW from potential phase transition at $\Tform$}

In addition to QCD phase transition, we also expect a possible phase transition at $\Tform$ leading to discrete $\mathbb{Z}_{N}$ symmetry breaking and the formations of domain walls. To describe the phase transition dynamics we need to know the complete UV model leading to the effective operators in Eq.~\eqref{eq:Leff}, as new particles and interactions modify the tree-level potential~\cite{Quiros:1999jp}. Depending on those UV model parameters, the thermal effective potential for the field $S$, $V(S,T)$, could allow nearly degenerate minima and thus a FOPT at a temperature of $\Tform \sim f \sim \sigma^{1/3}$. As discussed for the QCD case, we can then expect another source of GW's from this FOPT. In Fig.~\ref{fig:GW:spectroscopy}, we show the expected GW spectra for $T_{\rm form}=2 \times 10^{5} \, \rm GeV$, $(\alpha_{\rm GW},\beta_{\rm GW}/H(T_{\rm form})) = (0.5,10^{4}), (0.5,10^{5})$ and compare them with the sensitivities of GW experiments.

\subsection{Hints from pulsar time array experiments}
\label{sec:pulsar-data}
The current and future pulsar timing array experiments including NANOGrav~\cite{McLaughlin:2013ira, Brazier:2019mmu}, EPTA~\cite{Lentati:2015qwp}, PPTA~\cite{Hobbs:2013aka}, MeerTime~\cite{Bailes:2018azh}, CHIME~\cite{Ng:2017djg}, and SKA~\cite{Janssen:2014dka} are sensitive to gravitational waves with frequencies in the range $10^{-9}\,\rm{Hz}-10^{-7}\,\rm{Hz}$. Since collapsing domain walls with $\Tann \approx 100 \, \rm MeV$ lead to GW's with a peak frequency within the above range [see Eq.~\eqref{eq:fgwDW}], one can probe the QCD-collapsed domain-wall scenario with the PTAs. 
A few years back, three of the current pulsar timing array experiments NANOGrav~\cite{NANOGrav:2020bcs}, EPTA~\cite{Chen:2021rqp}, and PPTA~\cite{Goncharov:2021oub} had reported strong evidence for a common-spectrum red process across pulsars in their data. The result was also confirmed by the IPTA collaboration~\cite{Perera:2019sca}, combining the data sets of the three collaborations. While the common-red spectrum was observed in all of the data sets, the Hellings-Downs correlations~\cite{Hellings:1983fr} required to confirm the observation of GW background was lacking back then.

\begin{figure}[t!]
\centering
    \includegraphics[width=0.7\textwidth]{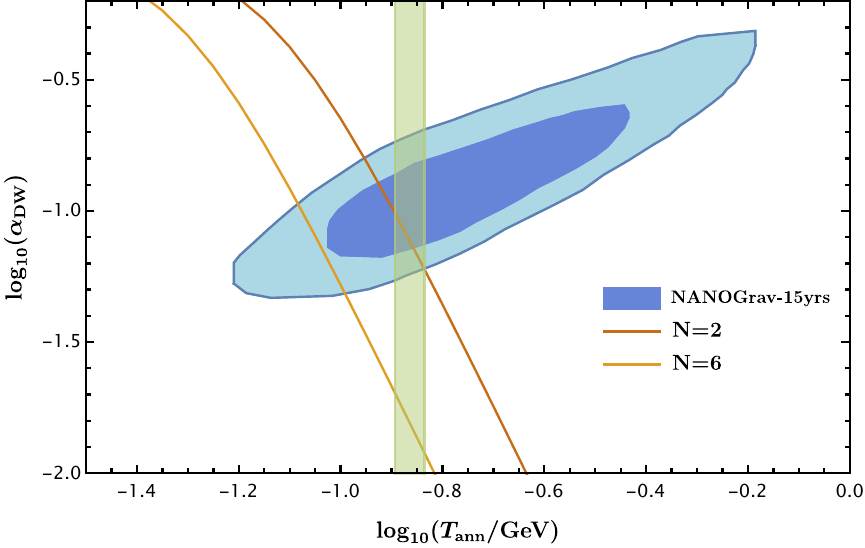}
    \caption{The preferred parameter space in $\Tann$ and $\alpha_{\rm DW}$ based on the NANOGrav 15-year data for the domain-wall model~\cite{NANOGrav:2023gor,NANOGrav:2023hvm}. The two contours correspond to one- and two-sigma confidence levels, respectively. The brown ($N=2$) and orange ($N=6$) curves are for the QCD-annihilated domain wall models with a relation between $\alpha_{\rm DW}$ and $\Tann$ after fixing $V_{\rm bias}$. The $N=2$ curve intersects with the two-sigma contour at points with the $\sigma$ values of $(0.66, 1.2)\times 10^{16}\,\mbox{GeV}^3$. For $N=6$, the intersections have $\sigma = (0.9, 1.3)\times 10^{15}\,\mbox{GeV}^3$. The green band indicates the range of the QCD phase transition temperatures for different $\theta$ angles (see Fig.~\ref{fig:phase:QCD} based on the LSM$q$ model).
    }
    \label{fig:DWnanograv}
\end{figure}

Recently, with a larger pulsar data set and longer experimental run-time, the Hellings-Downs correlation curve has been observed by NANOGrav at 3 sigma~\cite{NANOGrav:2023gor}, EPTA at 3 sigma~\cite{Antoniadis:2023ott}, PPTA at 2 sigma~\cite{Reardon:2023gzh} and CPTA at 4.6 sigma~\cite{Xu:2023wog}. These results suggest the first observation of SGWB. The domain wall collapse during the radiation-domination era as a new-physics explanation for the observed GW background has been considered by the NANOGrav collaboration~\cite{NANOGrav:2023hvm}. In  Fig.~\ref{fig:DWnanograv}, we quote NANOGrav's one- and two-sigma contours in the $\Tann$ and $\alpha_{\rm DW}$ plane that is favored by the NANOGrav 15-year data-set. Where $\alpha_{\rm DW}$ is defined as the fraction of domain-wall energy density over the critical energy density at the time of annihilation, which is given by $\alpha_{\rm DW} =\mathcal{F}/(1+\mathcal{F})$, where $\mathcal{F}$ is given by Eq.~\eqref{eq:Ffactor}. Using Eq.~\eqref{eq:Tann} and Eq.~\eqref{eq:Ffactor} with $V_{\rm bias}\approx (100 \, \rm{MeV})^{4}$ for $N=2$, and $V_{\rm bias}\approx (67 \,\rm {MeV})^{4}$ for $N=6$, one can obtain a relation between $\alpha_{\rm DW}$ and $\Tann$, which is shown in Fig.~\ref{fig:DWnanograv} by the brown ($N=2$) and orange ($N=6$) curves. For the $N=6$ case, we expect to have a sequential collapse of domain walls.  The GW spectrum shown in Fig.~\ref{fig:GW:Ann:XPT} is dominated by the smallest $T_{\rm ann}$ or smallest $V^{ij}_{\rm bias}$, which from Fig.~\ref{fig:Vij:dist} is given by $V_{\rm bias}^{1/4}\approx 67 \,\rm MeV$.

\subsection{GW spectroscopy}

\begin{figure}[t!]
\centering
    \includegraphics[width=0.7\textwidth]{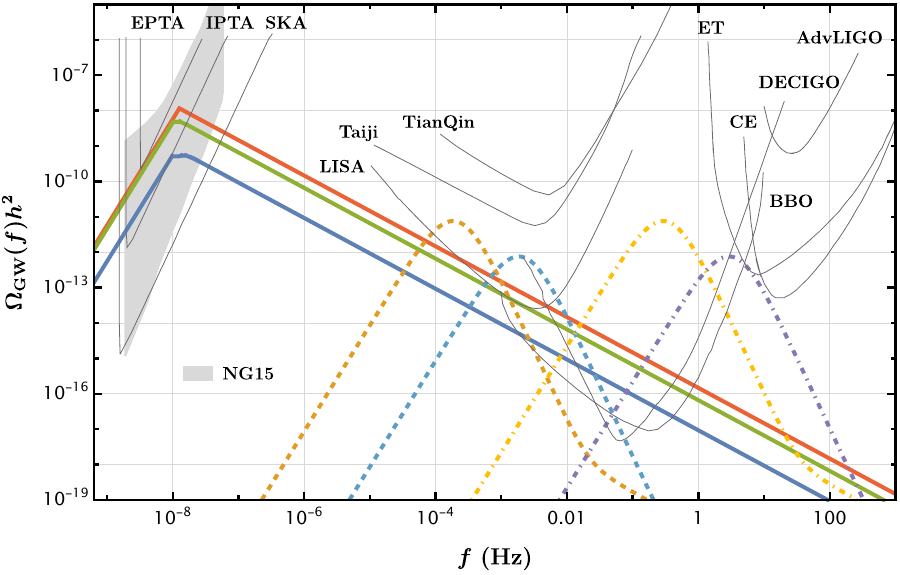}
    \caption{Gravitational wave spectroscopy in the QCD-collapsed domain wall models together with various existing and future experimental sensitivities. 
    Solid lines are from domain-wall annihilation with $(N=2,\sigma=1\times10^{16}~{\rm GeV^3})$ (red), $(N=4,\sigma=3\times10^{15}~{\rm GeV^3})$ (green) and $(N=6,\sigma=1\times10^{15}~{\rm GeV^3})$ (blue). Dashed lines are from QCD first-order phase transition with $\left(\alpha_{\rm GW}, \beta_{\rm GW}/H(T^{\rm QCD}_n)\right)=\left(0.5,10^4\right)$ (orange) and $\left(\alpha_{\rm GW},\beta_{\rm GW}/H(T^{\rm QCD}_n)\right)=\left(0.5,10^5\right)$ (blue). Dot-dashed lines are from possible first-order phase transition of the discrete symmetry breaking with  $\left(\alpha_{\rm GW},\beta_{\rm GW}/H(T_{\rm form})\right)=\left(0.5,10^4\right)$ (yellow) and  $\left(\alpha_{\rm GW},\beta_{\rm GW}/H(T_{\rm form})\right)=\left(0.5,10^5\right)$ (purple). The gray band denotes the rough range of GW spectrum observed by NANOGrav~\cite{NANOGrav:2023gor}.
    } 
    \label{fig:GW:spectroscopy}
\end{figure}

In Fig.~\ref{fig:GW:spectroscopy}, we show the predicted GW spectra from domain-wall annihilation, QCD and discrete-symmetry breaking phase transitions, assuming that both QCD and discrete-symmetry phase transitions are first-order. The detailed formulas for GW spectra from phase transition can be found in Appendix~\ref{sec:PTfomula}. The GW spectroscopy based on the QCD-anomalous discrete symmetries span more than 10 orders of magnitude in frequencies from $10^{-9}$ Hz to 100 Hz. For comparison, we also show the sensitivities from the current and future GW experiments: ET~\cite{Maggiore:2019uih}, AdvLIGO~\cite{LIGOScientific:2016jlg}, DECIGO~\cite{Kawamura:2011zz}, TianQin~\cite{TianQin:2015yph}, Taiji~\cite{Ruan:2018tsw}, LISA~\cite{2017arXiv170200786A}, SKA~\cite{5136190}, IPTA~\cite{Manchester:2013ndt},
EPTA~\cite{Kramer:2013kea}, CE~\cite{Reitze:2019iox}, BBO~\cite{Crowder:2005nr}. In this plot, the preferred GW spectrum range from NANOGrav results is shown in the gray band.

\section{Discussion and conclusions}
\label{sec:discussion}

In our setup, we have assumed $\theta_0 = 0$ in Eq.~\eqref{eq:Leff}, which is related to the strong CP problem. This could be realized within Nelson-Barr like models to solve the strong CP problem~\cite{Nelson:1983zb,Nelson:1984hg,Barr:1984qx}. Here, we provide a simple and explicit Nelson-Barr-like model based on the two-Higgs-doublet model to justify our choice of $\theta_0 = 0$ (see Ref.~\cite{Preskill:1991kd} for related discussion without solving the strong CP problem). Introducing one vector-like fermion $\psi_{L, R}$ with the same SM gauge charges as $u_R$ in the up-quark-sector, one has the CP-conserving and renormalizable Lagrangian as
\beqa
\label{eq:NS-lag}
\mathcal{L} &\supset& Y^u_{ij} \, \widetilde{H}_u \, \overline{Q}_{iL}\,u_{jR} + (\eta_j\,\phi + \kappa_j\,\phi^*)\,\bar{\psi}_L \, u_{jR} + \mu\,\bar{\psi}_L \, \psi_R \nonumber \\
&&\,+\,Y^d_{ij} \, H_d \, \overline{Q}_{iL}\,d_{jR} 
\,+\,\mbox{h.c.}\,-\, V(H_u, H_d, \phi) ~.
\eeqa
Here, $i, j = 1, 2, 3$ are family indices. $\widetilde{H}_u \equiv i \sigma_2 H^*_u$ with $H_u$ as a weak doublet. With real numbers for the couplings $Y_u^{ij}$, $Y_d^{ij}$, $\eta_j$ and $\kappa_j$ and vanishing $\theta$ angle at the UV scale, this Lagrangian is invariant under CP as well as the following ``Nelson-Barr discrete symmetry"
\beqa
\mathbb{Z}^{\rm NB}_2:&& H_u \rightarrow - H_u\,,\quad H_d \rightarrow H_d\,,\quad
\phi \rightarrow - \phi\,, \nonumber \\
&&u_R \rightarrow -u_R\,, \quad \psi_{L, R} \rightarrow \psi_{L, R}\,, \quad Q_L \rightarrow Q_L\,, \,\quad d_R \rightarrow d_R  ~. 
\eeqa
From minimizing the scalar potential, both CP and $\mathbb{Z}^{\rm NB}_2$ are spontaneously broken by the scalar vacuum expectation values with four degenerate vacua: $(\langle H_d \rangle, \langle H_u \rangle, \langle \phi \rangle)=$$(v\,\cos{\beta}, v\,\sin{\beta}, f\,e^{i\alpha})$, $(v\,\cos{\beta}, v\,\sin{\beta}, f\,e^{-i\alpha})$, $(v\,\cos{\beta}, -v\,\sin{\beta}, -f\,e^{i\alpha})$, $(v\,\cos{\beta}, -v\,\sin{\beta}, -f\,e^{-i\alpha})$ with $v = 246$~GeV. Note that the first and second vacua are related to the third and fourth ones by $\mathbb{Z}^{\rm NB}_2$, while the first and third ones are related to the second and fourth ones by CP. The Dirac CP violating phase in the Cabibbo-Kobayashi-Maskawa (CKM) matrix is related to $\alpha$ and other model parameters in Eq.~\eqref{eq:NS-lag}, while the strong CP violating phase is $\theta = \mbox{arg}[\mbox{det}(Y^u \langle \widetilde{H}_u \rangle)] + \mbox{arg}[Y^d]$, which equals to $0$ or $\pi$ and depend on the vacuum domains of $\mbox{arg}(Y^u \langle H_u \rangle)$. For instance, if $\mbox{arg}[\mbox{det}(Y^u)] = \mbox{arg}[\mbox{det}(Y^d)] = 0$, one has $\theta_0 = 0$ and $\theta = \mbox{arg}[\langle H_u \rangle] = 0$ or $\pi$. Furthermore, because there are only three right-handed fermions $u^i_R$ that are odd under $\mathbb{Z}^{\rm NB}_2$, this discrete symmetry $\mathbb{Z}^{\rm NB}_2$ is anomalous under QCD. So, this simple model serves as an example of QCD-anomalous discrete symmetries, which plays an essential role to solve the strong CP problem.~\footnote{We also note that the domain walls associated with spontaneously breaking of the CP symmetry are not collapsed by the QCD effects (similar to the axion domain wall problem~\cite{Sikivie:1982qv,Chang:1998tb,Sikivie:2006ni}). So, additional small and explicit CP-violation operators are needed to collapse the CP domain walls without generating a sizable strong CP phase. 
}

The domain-wall collapse, in addition to the production of gravitational waves, could also form primordial black holes (PBH)~\cite{Widrow:1989fe, Ferrer:2018uiu, Gelmini:2022nim, Gelmini:2023ngs}. 
The relevant quantity which determines whether PBHs are produced is $p(t)=2 G M_{\rm DW}(t)/L(t)$, the ratio of Schwarzschild radius of the domain wall to the size of the wall. When $p(t)\geq 1$, domain walls could collapse into black holes. Including both the vacuum energy and surface energy, the mass of the domain wall $M_{\rm DW}\approx \frac{4\pi}{3} 
 \, V_{\rm bias} L^{3} + 4 \pi \sigma L^{2}$. It is clear that a maximum of $p(t)$ is reached when $L(t)$ is maximum. As walls decelerate due to the potential bias term with a deceleration of $V_{\rm bias}/\sigma$, the maximum wall size is reached when the wall velocity becomes zero and the wall starts to shrink. The maximum wall size is estimated to be  
 \begin{equation}
    L_{\rm max} \simeq \frac{\left[ \mathcal{A} + \gamma(v_{0})v_{0}\right]}{\mathcal{A}} \frac{\sigma}{ V_{\rm bias}} \equiv \tilde{\delta}\,\frac{\sigma}{V_{\rm bias}} \, ,
 \end{equation}
where $v_{0}$ is the initial wall velocity. The maximum value of $p(t)$, $p_{\rm max}$, is then given by
\begin{equation}
 p_{\rm max} = \alpha_{\rm DW}\frac{\tilde{\delta}^{2}}{4 \mathcal{A}^{2}}\left( 1 + \frac{3}{ \tilde{\delta}}\right)  = 0.29 \, \frac{\alpha_{\rm DW}}{0.1} \left (\frac{\tilde{\delta}}{1.62} \right)^{2} \left(\frac{\mathcal{A}}{0.8} \right)^{-2} \frac{\left( 1 + 1.85 \frac{1.62}{\tilde{\delta}}\right)}{2.85} \, . 
\end{equation}
Here, we have used $v_{0}=0.42$ and $\mathcal{A}=0.8$, and $\alpha_{\rm DW}$ is the parameter used in the PTA result (see Fig.~\ref{fig:DWnanograv}). As long as $\alpha_{\rm DW}< 0.33$, we have $p_{\rm max}<1$ and PBH's are not produced from domain wall collapse. For the case with $\alpha_{\rm DW}> 0.33$, PBH's might be formed with their abundance determined by the details of the collapse process~\cite{Kawasaki:2014sqa} including sphericity~\cite{Widrow:1989fe}.

We also briefly comment on implications on quark nuggets for the QCD-collapsed domain wall scenario. The possibility of first-order phase transitions for some domains with a nonzero $\theta$ angle could fulfil the mechanism of ``Cosmic Separation of Phases", as pointed out in Ref.~\cite{Witten:1984rs}. Interestingly, collapsed domain walls can serve an alternative way to form quark nuggets. As the walls collapse, the baryon number is accumulated into a small pocket with the quark degenerate Fermi pressure to withhold the vacuum pressure. For both formation mechanisms, an initial nonzero baryon-number chemical potential is required to exist from early universe. The usual calculations of quark nugget properties are based on the $\theta = 0$ domain~\cite{Farhi:1984qu} (see also \cite{Holdom:2017gdc} for updates). The scenario studied in this paper could suggest other types of quark nuggets with a nontrivial QCD vacuum inside (the one with a nonzero $\theta$). For the phenomenological approach by introducing a ``bag parameter", one can define a generalized bag parameter: $B_{({\rm{had}})_{\theta=0}, (q/g)_{\theta}}$, which are related to the vacuum energy differences between the hadronic phase with $\theta=0$ and quark-gluon plasma phase with a generic $\theta$. Based on the LSM$q$ model, the bag parameter is calculated to be a monotonic decreasing function of $\theta$ from 0 to $\pi$ with $B_{({\rm{had}})_{\theta=0}, (q/g)_{\theta =0}} = (222\,\mbox{MeV})^4$ and $B_{({\rm{had}})_{\theta=0}, (q/g)_{\theta =\pi}} = (219\,\mbox{MeV})^4$. The stability of quark nuggets with a nonzero $\theta$ requires its mass per baryon to be lighter than the nucleon mass with the same $\theta$ (or even different $\theta$ if the quantum decay into nucleons with a different $\theta$ angle is allowed; Ref.~\cite{Lee:2020tmi} has suggested that the nucleon mass decreases monotonically as $\theta$ increases). Quark nuggets with a nonzero $\theta$ could be more stable than the ordinary quark nuggets because the nontrivial topological angle can further protect them from destructions. The properties, formation and evolution of this new type of quark nuggets deserve further investigation.

In this paper, we have considered only the case with the discrete-symmetry-breaking scale $f$ after inflation or reheating. For the opposite case, our universe could end up with a QCD with $\theta\neq 0$ without a domain wall in the visible universe. Since the current universe is consistent with a QCD with (approximately) zero $\theta$, quantum or thermal tunneling are needed to transit the vacuum to the QCD with zero $\theta$. For a high scale $f$ much above the QCD scale $\sim 100$~MeV, those tunneling rates are exponentially suppressed and a viable scenario is hardly anticipated. For the case with $f$ close to the QCD scale, the tunneling rates can be large enough, although additional QCD-charged fermions to mediate the QCD-anomalous discrete symmetry could have a too low mass to be phenomenologically viable.   


The scenario with QCD-anomalous discrete symmetries has the interesting interplay between the effective $\theta$ angle and QCD phase transition. One may wonder whether a similar interplay exists for QCD axion models, where different Hubble patches could have different axion field values or also different effective $\theta$ angles. The existence of an interesting interplay relies on the relation of the QCD phase transition temperature $T_{\rm QCD}$ and the axion oscillation temperature $T_{\rm osc}$, when the axion particle mass is comparable to the Hubble scale and the axion field starts to oscillate around the effective $\theta = 0$ value. Based on the dilute instanton gas calculation~\cite{Turner:1985si,DiLuzio:2020wdo}, the oscillation temperature $T_{\rm osc}$ is qualitatively higher than $T_{\rm QCD}$. Effectively, one has a zero $\theta$ during the QCD phase transition time with a crossover phase transition. Some interesting consequences for axion properties based on first-order QCD phase transitions (studied in Refs.~\cite{DeGrand:1985uq,Hindmarsh:1991ay,Kim:2018knr}) may not happen.

In conclusion, we have demonstrated that the PTA-data preferred domain-wall models could have an interesting interplay with the QCD dynamics with a nonzero $\theta$ angle, assuming an underlying QCD-anomalous discrete symmetry. The discrete-symmetry-breaking scale is around 100~TeV with the domain-wall annihilation temperature of around 100 MeV. Some domains with an effective large $\theta$ angle could undergo a first-order QCD phase transition, which can lead to other phenomenological consequences including GW's at a higher frequency. Many future GW experiments could test the QCD-collapsed domain wall models studied here.

\vspace{1cm}
\subsubsection*{Acknowledgments}
We thank Andrew Long, Pedro Schwaller, Luca Visinelli, Carlos Wagner, Chen Zhang and Ariel Zhitnitsky for useful discussion. The work is supported by the U.S. Department of Energy under the contract DE-SC-0017647. YB is grateful to the Mainz Institute for Theoretical Physics (MITP) of the Cluster of Excellence PRISMA+ (Project ID 39083149), where this work was initialized. This work was completed at the Aspen Center for Physics, which is supported by National Science Foundation grant PHY-2210452.

\appendix
\section{QCD PT with a non-zero $\theta$ in the LSM$q$ model}
\label{sec:appendix}

In this appendix, we study the finite temperature QCD PT with a non-zero $\theta$ based on the 3-flavor LSM$q$ model. The mean-field approximation for the 3-flavor LSM$q$ is studied in~\cite{Lenaghan:2000ey,Schaefer:2008hk,Mitter:2013fxa}. In this model, the field $\Phi$ which transforms as a bi-fundamental under $U(3)_L\times U(3)_R$ is parameterized as 
\begin{equation}\label{eq:LSMq:Phi}
\Phi = T_a\left(\sigma_a+i\pi_a\right) \, ,
\end{equation}
where $T_a=\Lambda_a/2$ with $a=0,\cdots,8$ are the nine generators of the $U(3)$ group with $\mbox{Tr}(T_a T_b) = \frac{1}{2}\delta_{ab}$. $\Lambda_{1-8}$ are the usual Gell-Mann matrices and $\Lambda_0=\sqrt{\frac{2}{3}}\mathbbm{I}_3$. The LSM$q$ potential is given by
\begin{equation}\label{eq:LSMq:V}
\begin{aligned}
	V(\Phi) &= \mu^2\Tr\left(\Phi^\dagger\Phi\right) + \lambda_1\left[\Tr\left(\Phi^\dagger\Phi\right)\right]^2 + \lambda_2\Tr\left[\left(\Phi^\dagger\Phi\right)^2\right] \\
	&\quad - \frac{\kappa}{2}\left[e^{-i\theta}\det\left(\Phi\right) + e^{i\theta}\det\left(\Phi^\dagger\right)\right] - \Tr\left[H\left(\Phi+\Phi^\dagger\right)\right] \, .
\end{aligned}
\end{equation}
Here, the first three terms are invariant under $U(3)_L\times U(3)_R$; the fourth term known as the effective 't~Hooft determinant models the $U(1)_A$ anomaly; the fifth term, which is the quark mass term, explicitly breaks both $SU(3)_L\times SU(3)_R$ and $U(1)_A$. In the last term, $H$ can be parametrized as $H = T_a h_a$.
Since this term explicitly breaks the chiral symmetry and thus carries the quantum numbers of the vacuum, only the diagonal entries, \ie, $h_0$, $h_3$, and $h_8$ are non-zero. Assuming that the isospin symmetry is intact, we set $h_3=0$ and reparametrize the $(0,8)$-basis into the (non-strange, strange) $(x,y)$ basis
\begin{equation}\label{eq:LSMq:basis}
	\begin{pmatrix}
		*_x \\
		*_y
	\end{pmatrix} = \frac{1}{\sqrt{3}}\begin{pmatrix}
		\sqrt{2} & 1 \\
		1 & -\sqrt{2}
	\end{pmatrix}\begin{pmatrix}
		*_0 \\
		*_8
	\end{pmatrix} ,\, *=\sigma,\pi,h \, .
\end{equation}

Assuming that only the condensates $\langle\sigma_{x,y}\rangle$ and $\langle\pi_{x,y}\rangle$ are non-zero, the vacuum potential is given by
\begin{equation}\label{eq:LSMq:Vvac}
\begin{aligned}
	V_{\rm vac} &= \frac{\mu^2}{2}\left(\langle\sigma_x\rangle^2+\langle\sigma_y\rangle^2+\langle\pi_x\rangle^2+\langle\pi_y\rangle^2\right) + \frac{\lambda_1}{4}\left(\langle\sigma_x\rangle^2+\langle\sigma_y\rangle^2+\langle\pi_x\rangle^2+\langle\pi_y\rangle^2\right)^2 \\
	&\quad + \frac{\lambda_2}{8}\left[\left(\langle\sigma_x\rangle^2+\langle\pi_x\rangle^2\right)^2+2\left(\langle\sigma_y\rangle^2+\langle\pi_y\rangle^2\right)^2\right] - h_x\langle\sigma_x\rangle - h_y\langle\sigma_y\rangle \\
	&\quad + \frac{\kappa}{4\sqrt{2}}\left[\left(2\langle\sigma_x\rangle\langle\pi_x\rangle\langle\pi_y\rangle-\langle\sigma_x\rangle^2\langle\sigma_y\rangle+\langle\sigma_y\rangle
\langle\pi_x\rangle^2\right)\cos\theta \right. \\
	&\quad\qquad\qquad   \left. -\left(2\langle\sigma_x\rangle\langle\sigma_y\rangle\langle\pi_x\rangle+\langle\sigma_x\rangle^2\langle\pi_y\rangle-\langle\pi_x\rangle^2\langle\pi_y\rangle\right)\sin\theta\right] \, .
\end{aligned}
\end{equation}
The parameters of the potential will be fixed by the observed meson masses and decay constants in the global QCD vacuum [$\langle\pi_{x,y}\rangle=0$ and $\theta=0$]. 

We define
\begin{equation}\label{eq:LSMq:mass-squared}
	m^2_{\phi_i,\phi_j} = \left.\frac{\partial^2V(\Phi)}{\partial\phi_i\partial\phi_j}\right\vert_{\rm global\,vacuum} ,\, \phi=\sigma,\pi;\, i,j=0,\cdots, 8 \,.
\end{equation}
Because CP is conserved in the global QCD vacuum, scalar and pseudoscalar modes are separated. We first list the scalar meson masses:
\begin{description}
	\item[$\bullet~a0(980)$:] \begin{equation}\label{eq:LSMq:masq}
		m_\alpha^2 = m^2_{\sigma_{1-3},\sigma_{1-3}} = \mu^2 + \lambda_1\left(\langle\sigma_x\rangle^2+\langle\sigma_y\rangle^2\right) + \frac{3}{2}\lambda_2\langle\sigma_x\rangle^2 + \frac{\kappa}{2\sqrt{2}}\langle\sigma_y\rangle \, .
	\end{equation}
	
	\item[$\bullet~K^*(1410)$:] \begin{equation}\label{eq:LSMq:mkasq}
		\begin{aligned}
		m_\kappa^2 = m^2_{\sigma_{4-7},\sigma_{4-7}} &= \mu^2 + \lambda_1\left(\langle\sigma_x\rangle^2+\langle\sigma_y\rangle^2\right) \\
		&\quad + \frac{\lambda_2}{2}\left(\langle\sigma_x\rangle^2+\sqrt{2}\langle\sigma_x\rangle\langle\sigma_y\rangle+2\langle\sigma_y\rangle^2\right) + \frac{\kappa}{4}\langle\sigma_x\rangle \, .
	\end{aligned}
	\end{equation}
	
	\item[$\bullet~f0(500),f0(1370)$:] These two singlet states will mix according to the mass-squared matrix
	\begin{equation}
		M_S^2 = \begin{pmatrix}
			m^2_{\sigma_0,\sigma_0} & m^2_{\sigma_0,\sigma_8} \\
			m^2_{\sigma_8,\sigma_0} & m^2_{\sigma_8,\sigma_8}
		\end{pmatrix} \, ,
	\end{equation}
	where
	\begin{align}
		(M_S^2)_{00} &= \mu^2 + \frac{\lambda_1}{3}\left(7\langle\sigma_x\rangle^2+4\sqrt{2}\langle\sigma_x\rangle\langle\sigma_y\rangle+5\langle\sigma_y\rangle^2\right) \nonumber \\
		&\quad   + \lambda_2\left(\langle\sigma_x\rangle^2+\langle\sigma_y\rangle^2\right)- \frac{\kappa}{6}\left(2\langle\sigma_x\rangle+\sqrt{2}\langle\sigma_y\rangle\right) \, , \\
		(M_S^2)_{11} &= \mu^2 + \frac{\lambda_1}{3}\left(5\langle\sigma_x\rangle^2-4\sqrt{2}\langle\sigma_x\rangle\langle\sigma_y\rangle+7\langle\sigma_y\rangle^2\right)\nonumber \\
		&\quad  + \frac{\lambda_2}{2}\left(\langle\sigma_x\rangle^2+4\langle\sigma_y\rangle^2\right) + \frac{\kappa}{12}\left(4\langle\sigma_x\rangle-\sqrt{2}\langle\sigma_y\rangle\right) \, , \\
		(M_S^2)_{01} &= (M_S^2)_{10} = \frac{2}{3}\lambda_1\left(\sqrt{2}\langle\sigma_x\rangle^2-\langle\sigma_x\rangle\langle\sigma_y\rangle-\sqrt{2}\langle\sigma_y\rangle^2\right) \nonumber \\
		&\quad + \frac{\lambda_2}{\sqrt{2}}\left(\langle\sigma_x\rangle^2-2\langle\sigma_y\rangle^2\right) + \frac{\kappa}{12}\left(\sqrt{2}\langle\sigma_x\rangle-2\langle\sigma_y\rangle\right) \, ,
	\end{align}
the eigenvalues are 
	\begin{align}
		m_\sigma^2 &= \frac{(M_S^2)_{00}+(M_S^2)_{11}-\sqrt{((M_S^2)_{00}-(M_S^2)_{11})^2+4(M_S^2)_{01}^2}}{2} \, , \\
		m_f^2 &= \frac{(M_S^2)_{00}+(M_S^2)_{11}+\sqrt{((M_S^2)_{00}-(M_S^2)_{11})^2+4(M_S^2)_{01}^2}}{2} \, .
	\end{align}
\end{description}

Next, we list the pseudoscalar meson masses:
\begin{description}
	\item[$\bullet~\pi$:] \begin{equation}\label{eq:LSMq:mpisq}
		m_\pi^2 = m^2_{\pi_{1-3},\pi_{1-3}} = \mu^2 + \lambda_1\left(\langle\sigma_x\rangle^2+\langle\sigma_y\rangle^2\right) + \frac{\lambda_2}{2}\langle\sigma_x\rangle^2 - \frac{\kappa}{2\sqrt{2}}\langle\sigma_y\rangle \, .
	\end{equation}
	
	\item[$\bullet~K$:] \begin{equation}\label{eq:LSMq:mKsq}
		\begin{aligned}
		m_K^2 = m^2_{\pi_{4-7},\pi_{4-7}} &= \mu^2 + \lambda_1\left(\langle\sigma_x\rangle^2+\langle\sigma_y\rangle^2\right) \\
		&\quad + \frac{\lambda_2}{2}\left(\langle\sigma_x\rangle^2-\sqrt{2}\langle\sigma_x\rangle\langle\sigma_y\rangle+2\langle\sigma_y\rangle^2\right) - \frac{\kappa}{4}\langle\sigma_x\rangle \, .
	\end{aligned}
	\end{equation}
	
	\item[$\bullet~\eta,\eta^\prime$:] These two singlet states will mix according to the mass-squared matrix
	\begin{equation}
		M_P^2 = \begin{pmatrix}
			m^2_{\pi_0,\pi_0} & m^2_{\pi_0,\pi_8} \\
			m^2_{\pi_8,\pi_0} & m^2_{\pi_8,\pi_8}
		\end{pmatrix} \, ,
	\end{equation}
	where
	\begin{align}
		(M_P^2)_{00} &= \mu^2 + \lambda_1\left(\langle\sigma_x\rangle^2+\langle\sigma_y\rangle^2\right) + \frac{\lambda_2}{3}\left(\langle\sigma_x\rangle^2+\langle\sigma_y\rangle^2\right) + \frac{\kappa}{6}\left(2\langle\sigma_x\rangle+\sqrt{2}\langle\sigma_y\rangle\right) \, , \\
		(M_P^2)_{11} &= \mu^2 + \lambda_1\left(\langle\sigma_x\rangle^2+\langle\sigma_y\rangle^2\right) + \frac{\lambda_2}{6}\left(\langle\sigma_x\rangle^2+4\langle\sigma_y\rangle^2\right) - \frac{\kappa}{12}\left(4\langle\sigma_x\rangle-\sqrt{2}\langle\sigma_y\rangle\right) \, , \\
		(M_P^2)_{01} &= (M_P^2)_{10} = \frac{\lambda_2}{3\sqrt{2}}\left(\langle\sigma_x\rangle^2-2\langle\sigma_y\rangle^2\right) - \frac{\kappa}{12}\left(\sqrt{2}\langle\sigma_x\rangle-2\langle\sigma_y\rangle\right) \, ,
	\end{align}
the eigenvalues are
	\begin{align}
		m_\eta^2 &= \frac{(M_P^2)_{00}+(M_P^2)_{11}-\sqrt{((M_P^2)_{00}-(M_P^2)_{11})^2+4(M_P^2)_{01}^2}}{2} \, , \\
		m_{\eta^\prime}^2 &= \frac{(M_P^2)_{00}+(M_P^2)_{11}+\sqrt{((M_P^2)_{00}-(M_P^2)_{11})^2+4(M_P^2)_{01}^2}}{2} \, .
	\end{align}
\end{description}

Using the partially conserved axial current relation, $\langle\sigma_{x,y}\rangle$ and $h_{x,y}$ are given by~\cite{Lenaghan:2000ey}
\begin{gather}\label{eq:LSMq:PCAC}
	\langle\sigma_x\rangle = f_\pi ,\, \langle\sigma_y\rangle = \frac{2f_K-f_\pi}{\sqrt{2}} \, , \\
	h_x = f_\pi m_\pi^2 ,\, h_y = \sqrt{2}f_Km_K^2 - \frac{f_\pi m_\pi^2}{\sqrt{2}} \, ,
\end{gather}
where $f_{\pi,K}$ are the pion and kaon decay constants, respectively.

Once we have the relation between potential parameters and experimental observables, we can fix the parameters using the observed values. We choose experimental values of $f_\pi$, $f_K$, $m_\pi$, $m_K$, $m_\eta^2+m_{\eta^\prime}^2$, and $m_\sigma=600$~MeV~\cite{ParticleDataGroup:2022pth} to fix the six potential parameters $\lambda_{1,2}, h_{x,y}, \mu^{2}, \kappa$. In this way the remaining meson masses, such as the masses of $\eta$ and $\eta^\prime$, could be checked against the measured values, and are in fact found to match with the experimental values.       

Once fixing the potential parameters, one can look at the vacuum structure of the potential at $T=0$ for different $\theta$. One has 
\begin{itemize}
	\item $(\theta,\langle\sigma_x\rangle,\langle\sigma_y\rangle,\langle\pi_x\rangle,\langle\pi_y\rangle)=(0,92,90,0.0,0.0)~\rm MeV$\quad (global vacuum) .
	\item $(\theta,\langle\sigma_x\rangle,\langle\sigma_y\rangle,\langle\pi_x\rangle,\langle\pi_y\rangle)=(0,-81,88,0.0,0.0)~\rm MeV$ .
	\item $(\theta,\langle\sigma_x\rangle,\langle\sigma_y\rangle,\langle\pi_x\rangle,\langle\pi_y\rangle)=(\pi,3.8,89,87,2.0)~\rm MeV$ .
	\item $(\theta,\langle\sigma_x\rangle,\langle\sigma_y\rangle,\langle\pi_x\rangle,\langle\pi_y\rangle)=(\pi,3.8,89,-87,-2.0)~\rm MeV$ .
\end{itemize}
Here, the first two vacua with $\theta=0$ are CP conserving with the first one being the global vacuum. The last two vacua with $\theta=\pi$ are degenerate and CP-violating. 

Next we want to derive the vacuum structure at finite temperature to determine the nature of finite-temperature QCD phase transition. In the LSM$q$~\cite{Pisarski:1996ne}, in addition to the potential $V_{\rm vac}$ in Eq.~\eqref{eq:LSMq:Vvac}, one also has the Yukawa interactions between the quarks and the meson fields given by
\begin{equation}\label{eq:LSMq:Yukawa}
	\mathcal{L}_{\rm Yukawa} = \overline{q}\left[i\slashed{\partial} - gT_a\left(\sigma_a+i\gamma^5\pi_a\right)\right]q ,\quad q=u,d,s \, .
\end{equation}
In fact at non-zero temperature the quark degrees of freedom are the only one giving rise the thermal potential
\begin{equation}\label{eq:LSMq:thermal}
	V_T = -\nu_q\int\frac{d^3p}{(2\pi)^3} T\log\left[1+e^{-\frac{E_q}{T}}\right] - \nu_s\int\frac{d^3p}{(2\pi)^3} T\log\left[1+e^{-\frac{E_s}{T}}\right] \,,
\end{equation}
where $\nu_q=24$, $\nu_s=12$, $E_q=\sqrt{p^2+M_q^2}$, and $E_s=\sqrt{p^2+M_s^2}$, with $M_q=g\sqrt{\langle\sigma_x\rangle^2+\langle\pi_x\rangle^2}/2$ and $M_s=g\sqrt{\langle\sigma_y\rangle^2+\langle\pi_y\rangle^2}/\sqrt{2}$. Because of the isospin symmetry, we require that the constituent light quark mass $M_q=gf_\pi/2$ should be roughly $1/3$ of a nucleon mass, which gives $g=6.6$ and accordingly the constituent $s$-quark mass $M_s \sim 422$~MeV. After fixing $V_T$, one has the total finite temperature potential as $V_T + V_{\rm vac}$. We can minimize this potential for any given $T$ and $\theta$ value to obtain the vacuum expectation values $\langle\sigma_{x,y}(T)\rangle, \langle\pi_{x,y}(T)\rangle$ as a function of temperature to study phase transition.

\begin{figure}[th!]
\centering
    \includegraphics[width=0.48\textwidth]{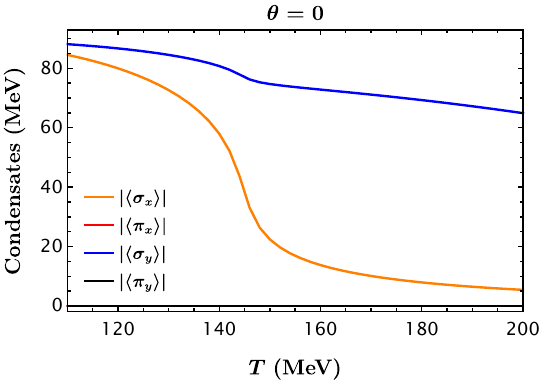}
    \includegraphics[width=0.48\textwidth]{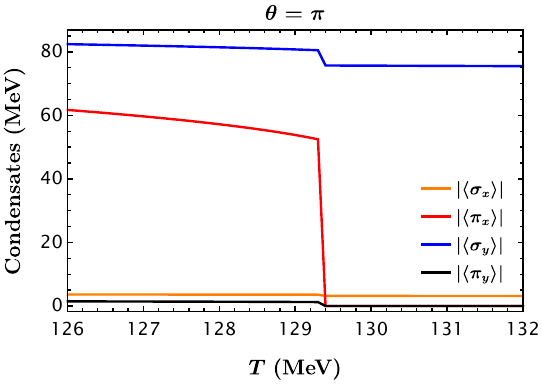}
    \caption{Thermal evolution of various condensates with $\theta=0$ (left) and $\theta=\pi$ (right). In the left panel one has a crossover phase transition at $T\approx146$~MeV, while in the right panel one has a first-order phase transition at $T\approx129$~MeV.
    }
    \label{fig:vev:T}
\end{figure}

In Fig.~\ref{fig:vev:T} we plot the thermal evolution of the condensates with $\theta=0,\pi$, respectively. One can see from the left panel that when $\theta=0$ there is a crossover phase transition at $T\approx146$~MeV as expected from the lattice QCD simulations. On the contrary, in the right panel with $\theta=\pi$ there is a first-order phase transition at $T\approx129$~MeV. 
The procedure can be carried out for different $\theta$ values to figure out the nature of phase transition. We use CosmoTransitions~\cite{Wainwright:2011kj} to identify the transitions and show the resulted $\theta-T$ phase diagram in Fig.~\ref{fig:phase:QCD}.

\section{Gravitational wave spectrum from phase transition}
\label{sec:PTfomula}

In this appendix we list the GW spectrum formulas for three processes: collision of bubbles, sound waves generated
from bubble expansion, and turbulence in plasma~\cite{Huber:2008hg,Caprini:2009fx,Caprini:2015zlo,Espinosa:2010hh,Hindmarsh:2015qta}. We also take into account the $\zeta$ factor which denotes the fraction of the universe undergoing FOPT. The spectrum is given by:
\allowdisplaybreaks
\begin{align}
   \Omega_{\rm col}(f)h^2 &\approx 1.4 \times 10^{-13}\, \frac{\zeta}{0.5}\, \left(\frac{\beta_{\rm GW}/H(T_n)}{10^{4}}\right)^{-2}\left(\frac{\kappa_{\rm col}\alpha_{\rm GW}}{1+\alpha_{\rm GW}}\right)^2\left(\frac{g_*(T_n)}{20}\right)^{-1/3} \nonumber \\
   &\times \left(\frac{0.11v_w^3}{0.42+v_w^2}\right)\frac{3.8(f/f_{\rm col})^{2.8}}{1+2.8(f/f_{\rm col})^{3.8}} \, ,  \\
   \Omega_{\rm sw}(f)h^2 &\approx 2.26 \times 10^{-14}\,\frac{\zeta}{0.5}\,\left(\frac{\beta_{\rm GW}/H(T_n)}{10^{4}}\right)^{-1}\left(\frac{\kappa_{\rm sw}\alpha_{\rm GW}}{1+\alpha_{\rm GW}}\right)^2\left(\frac{g_*(T_n)}{20}\right)^{-1/3}v_w \nonumber \\
   &\times \left(\frac{f}{f_{\rm sw}}\right)^3\left(\frac{7}{4+3(f/f_{\rm sw})^2}\right)^{7/2} \, , \\
   \Omega_{\rm turb}(f)h^2 &= 2.8\times 10^{-8}\,\frac{\zeta}{0.5}\,\left(\frac{\beta_{\rm GW}/H(T_n)}{10^{4}}\right)^{-1}\left(\frac{\kappa_{\rm turb}\alpha_{\rm GW}}{1+\alpha_{\rm GW}}\right)^{3/2}\left(\frac{g_*(T_n)}{20}\right)^{-1/3}v_w \nonumber \\
   &\times \frac{\left(\frac{f}{f_{\rm turb}}\right)^3}{\left(1+\frac{f}{f_{\rm turb}}\right)^{11/3}\left(1+\frac{8\pi f}{h^{*}}\right)} \, ,
\end{align}
where $v_w$ is the wall velocity and $\kappa_{\rm col,\,sw,\,turb}$ denotes the fraction of vacuum energy that is converted into kinetic energy, bulk motion of the fluid, and of the turbulence, respectively. They are given by \begin{align}
	\kappa_{\rm col} &= \frac{1}{1+0.715 \, \alpha_{\rm GW}}\left(0.715 \, \alpha_{\rm GW}+\frac{4}{27}\sqrt{\frac{3 \, \alpha_{\rm GW}}{2}}\right) \,, \\
	\kappa_{\rm sw} &= \frac{\alpha_{\rm GW}}{0.73+0.082\, \sqrt{\alpha_{\rm GW}}+\alpha_{\rm GW}} \,, \\
	\kappa_{\rm turb} &= \xi_{\rm turb}\kappa_{\rm sw} \,,
\end{align} 
where $\xi_{\rm turb}\sim 0.1$ is the fraction of turbulent bulk motion. The red-shifted peak frequencies of GW spectra are 
\begin{align}
	f_{\rm col} &\approx 1.6\times 10^{-6}~{\rm Hz}\times\left(\frac{0.62}{1.8-0.1v_w+v_w^2}\right)\left(\frac{\beta_{\rm GW}/H(T_n)}{10^{4}}\right)\left(\frac{T_n}{125~{\rm MeV}}\right)\left(\frac{g_*(T_n)}{20}\right)^{1/6} \,, \\
	f_{\rm sw} &\approx 1.8\times10^{-4}~{\rm Hz}\times\frac{1}{v_w}\left(\frac{\beta_{\rm GW}/H(T_n)}{10^{4}}\right)\left(\frac{T_n}{125~{\rm MeV}}\right)\left(\frac{g_*(T_n)}{20}\right)^{1/6} \,, \\
	f_{\rm turb} &\approx 2\times10^{-4}~{\rm Hz}\times\frac{1}{v_w}\left(\frac{\beta_{\rm GW}/H(T_n)}{10^{4}}\right)\left(\frac{T_n}{125~{\rm MeV}}\right)\left(\frac{g_*(T_n)}{20}\right)^{1/6} \,.
\end{align}
The red-shifted Hubble parameter to today is given by
\begin{equation}
    h_{*} \approx 1.6 \times 10^{-8}~{\rm Hz}\left(\frac{T_n}{125~{\rm MeV}}\right)\left(\frac{g_*(T_n)}{20}\right)^{1/6} \, .
\end{equation}

\setlength{\bibsep}{6pt}
\providecommand{\href}[2]{#2}\begingroup\raggedright\endgroup

\end{document}